\documentclass[twocolumn,floatfix]{revtex4}
\usepackage{graphicx}
\begin{document}\title{Amorphous Systems in Athermal, Quasistatic Shear}
\author{Craig E. Maloney$^{(1,2)}$}
\altaffiliation{Current Address: KITP, University of California, Santa Barbara, California 93106, U.S.A.}
\author{Ana\"el Lema\^{\i}tre$^{(1,3)}$}
\affiliation{
$^{(1)}$ Department of physics, University of California, Santa Barbara, California 93106, U.S.A.}
\affiliation{$^{(2)}$ Lawrence Livermore National Lab - CMS/MSTD, Livermore, California 94550, U.S.A.}
\affiliation{$^{(3)}$ Institut Navier -- LMSGC, 2 all\'ee K\'epler, 77420 Camps-sur-Marne, France}
\date{\today}

\begin{abstract}
We present results on a series of 2D atomistic computer simulations of amorphous systems subjected to simple shear in the athermal, quasistatic limit.
The athermal quasistatic trajectories are shown to separate into smooth, reversible elastic branches which are intermittently broken by discrete catastrophic  plastic events.
The onset of a typical plastic event is studied with precision, and it is shown that the mode of the system which is responsible for the loss of stability has structure in real space which is consistent with a quadrupolar source acting on an elastic matrix.
The plastic events themselves are shown to be composed of localized shear transformations which organize into lines of slip which span the length of the simulation cell, and a mechanism for the organization is discussed.
Although within a single event there are strong spatial correlations in the deformation, we find little correlation from one event to the next, and these transient lines of slip are not to be confounded with the persistent regions of localized shear --- so-called "shear bands" --- found in related studies.
The slip lines gives rise to particular scalings with system length of various measures of event size.
Strikingly, data obtained using three differing interaction potentials can be brought into quantitative agreement after a simple rescaling, emphasizing the insensitivity of the emergent plastic behavior in these disordered systems to the precise details of the underlying interactions.
The results should be relevant to understanding plastic deformation in systems such as metallic glasses well below their glass temperature, soft glassy systems (such as dense emulsions), or compressed granular materials.

\end{abstract}
\maketitle

\section{Introduction}
\label{seq:intro}

In crystalline materials it is generally accepted that the microstructural objects which govern deformation and flow are a class of topological defects known as dislocations. Most work in the field of crystalline plasticity focuses on describing deformation in terms of the underlying dislocation dynamics.
In the case of non-crystalline systems the situation is not so clear, even though 
a broad category of systems--including metallic glasses; clays and soils; pastes, 
foams, gels, and other so-called ``soft'' materials--seem to share a few hallmark traits.
In the past several decades much work regarding the underlying microscopic processes of amorphous plastic flow, has left many unanswered questions and much controversy.
Perhaps the most important of these questions is whether or not there exist some sort of microstructural defects in the materials which, roughly speaking, play the role of the dislocations in crystals~\cite{Gilman1975,Gilman1975a,Chaudhari1979,Chaudhari1982,Shi1983}.

Some of the earliest computer simulations of metallic glasses by Maeda and Takeuchi and co-workers~\cite{Kobayashi1980,Maeda1981} and experiments on rafts of soap bubbles by Argon and Kuo~\cite{Argon1979a} and Argon and Shi~\cite{Argon1982} revealed that the atomic motions during plastic shear were confined to clusters of particles with a size of several particle diameters across.
These observations inspired Argon to propose a theoretical scheme based on a mean-field treatment of transitions in local regions of space, ``Shear Transformation Zones'' (STZs), where each transformation contributes a quantum of plastic shear strain~\cite{Argon1979}.
Drawing on Argon's works, Falk and Langer~\cite{Falk1998}, introduced mean-field equations of motion for the number density of STZs and showed that such a theory could account qualitatively for many aspects of the phenomenology of sheared metallic glasses such as strain hardening, the Bauschinger effect, and the emergence of a yield stress. 
The basic picture of an STZ may not be limited in utility to metallic glasses and could play an important part in the physics of other amorphous materials under shear such as foams, pastes, granular materials and the like~\cite{Debregeas2001,Lauridsen2002,Lauridsen2004,Pratt2003,Bonn2002,Coussot2002,Howell1999a,Lemaitre2002a,Lemaitre2002b}.

These initial treatments all neglected spatial interactions of STZs. However, one might expect that the rearrangement of an STZ should induce quadrupolar elastic displacements at long range, in analogy with the transformation of an Eshelby inclusion,~\cite{Eshelby1957,Picard2005} or the nucleation of a dislocation loop. Schematic, mesoscopic models constructed to account for such elastic interactions, first proposed by Bulatov and Argon, and later extended by others,~\cite{Bulatov1994,Bulatov1994a,Bulatov1994b,Langer2001,Baret2002,Onuki2003,Picard2005} have been shown to predict various sorts of localization of deformation. This mechanism might provide a physical explanation for an important technological problem: shear-banding, that is the localization of deformation to narrow bands, which is observed in many systems, including: metallic glasses~\cite{Wright2001,Wright2001a,Schuh2003a}, sheared rafts of bubbles~\cite{Argon1982}, sheared foams confined between glass plates~\cite{el1999}, dry foams~\cite{Kabla2003}, and granular materials~\cite{Aharonov2002,Kuhn1999,Kuhn2003}.

Despite these successes, theories of plasticity in amorphous materials remain controversial because they rely on many assumptions which are difficult to check in precise ways.  In particular, these elusive STZs --unlike dislocations--cannot be identified \emph{a priori} as any sort of topological defect~\footnote{The observation of Eshelby-like transitions in dry foams~\cite{Kabla2003} is an exception since in these systems, plasticity involves elementary T1 events, which \emph{are} associated with topological defects}. The athermal, quasi-static simulations we present here provide us a starting point to perform such observations since they allow for the isolation and identification of elementary transitions between mechanically stable states~\cite{Malandro1998,Malandro1999}. On these grounds, one might expect that an individual, elementary, quadrupolar, Eshelby-like rearrangement might be associated with any particular single transition between mechanically stable states, but as we will see, the reality is more complex. Our work is a first attempt to simultaneously identify elementary shear induced rearrangements in simulations of amorphous solids from both the perspective of the energy landscape and real-space.



This paper is organized as follows.
We first, in section~\ref{sec:background}, outline our athermal, quasistatic (AQS) algorithm, and provide a physical rationale for its use, discussing the types of physical systems to which we expect our treatment to apply and its limitations.
Section~\ref{sec:elasticSegments} deals with the nature of the smooth elastic segments of AQS trajectories, and in particular how the trajectories break down at the onset of individual catastrophic events.
The nucleation of one particular plastic event in a moderately sized system will be used as a case study.
Section~\ref{sec:plasticEvent} deals with the nature of the individual plastic cascades themselves and their spatial structure.
Again, a typical event will be used as a case study.
Finally, in section~\ref{sec:scaling}, we will show how the nature of the plastic events discussed in section~\ref{sec:plasticEvent} dictates particular scalings with the system size of the stress and energy relaxation during the plastic events and the strain interval between successive events.
\section{The Athermal, Quasistatic Limit}
\label{sec:background}

\subsection{Timescales}

Athermal, quasi-static (AQS) simulations have been used in several recent studies~\cite{Malandro1998,Malandro1999,Utz2000,Schuh2003,Nandagopal2003,Bailey2004} of plasticity in amorphous solids. 
The AQS algorithm simply consists of repeated alternating steps of: 1) minimization of the potential energy of all the particles in the simulation cell~\footnote{The same procedure is used to define so-called Inherent Structures in conventional MD simulations~\cite{Stillinger1982,Stillinger1983}.} and 2) application of a small, homogeneous strain to all particles and simulation cell boundaries.
This simulation technique was introduced by Kobayashi, Maeda and Takeuchi~\cite{Maeda1978a,Kobayashi1980,Maeda1981} as a way to bypass intrisic limitations of MD simulations to reach long timescales, and therefore low shear rates. Such limitations remain even with modern computers.

The AQS algorithm relies on the idea that in the absence of external drive, amorphous solids remain close to a mechanically stable state in a complex potential energy landscape. For molecular or metallic glasses, this assumption is reasonable as soon as the bath temperature is low compared to the glass transition temperature $T_g$. A more precise bound can be obtained when these solids are submitted to some constant deformation rate $\dot\gamma$, considering that the thermal relaxation should be compared to $\dot\gamma$. The athermal limit correspond to the situation when $\dot\gamma>>1/\tau_{\rm relax}$, where
$\tau_{\rm relax}$ characterizes the thermally activated escape of the system from a local minimum.
In this limit, escape from local minima is primarily induced by strain and not by thermal activation~\cite{Yamamoto1998,Angelani2002}.
Because a low temperature limit is taken, this situation is likely to be relevant to different systems than metallic or molecular glasses. Foams, granular materials close to jamming, and several instances of soft glassy systems are intrinsically athermal and would likewise remain in a mechanically stable configuration in the absence of any external drive. 

\begin{figure}
\includegraphics[width=.45\textwidth]{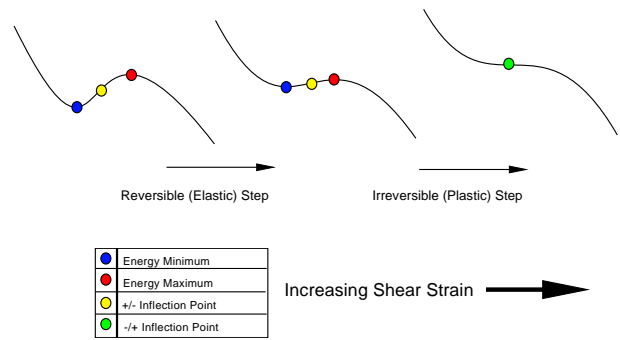}
\caption{A schematic representation of deformation-induced changes of a local minimum in the potential energy landscape. The shape of the landscape varies continuously as the strain is increased going from left to right with both the location and height of the minimum changing.}
\label{fig:landscapeCartoon}
\end{figure}

When these amorphous solids are submitted to small amounts of deformation, they smoothly follow deformation-induced continuous changes of a local minimum. This process is illustrated in figure~\ref{fig:landscapeCartoon}. Strain induces a bias on the potential energy landscape, and the material configuration tracks the location of a single energy minimum as it moves smoothly through configuration space. This kind of motion is completely reversible in that if we reverse the sense of the imposed strain, the system returns to its original configuration. As we will see, it is possible to solve analytically for the trajectories of the system during this smooth motion, and these trajectories determine the elastic constants of the material.~\cite{Wittmer2002,Tanguy2002,Lemaitre2005} Of course, reversibility holds only for small enough amounts of strain such that the minimum remains stable. For increasing strains, this smooth behavior must eventually break down, as the energy minimum in which the system resides flattens out and collides with a saddle point~\cite{Malandro1998,Malandro1999,Lacks2001,Maloney2004a}.

\begin{figure}
  \includegraphics[width=.45\textwidth]{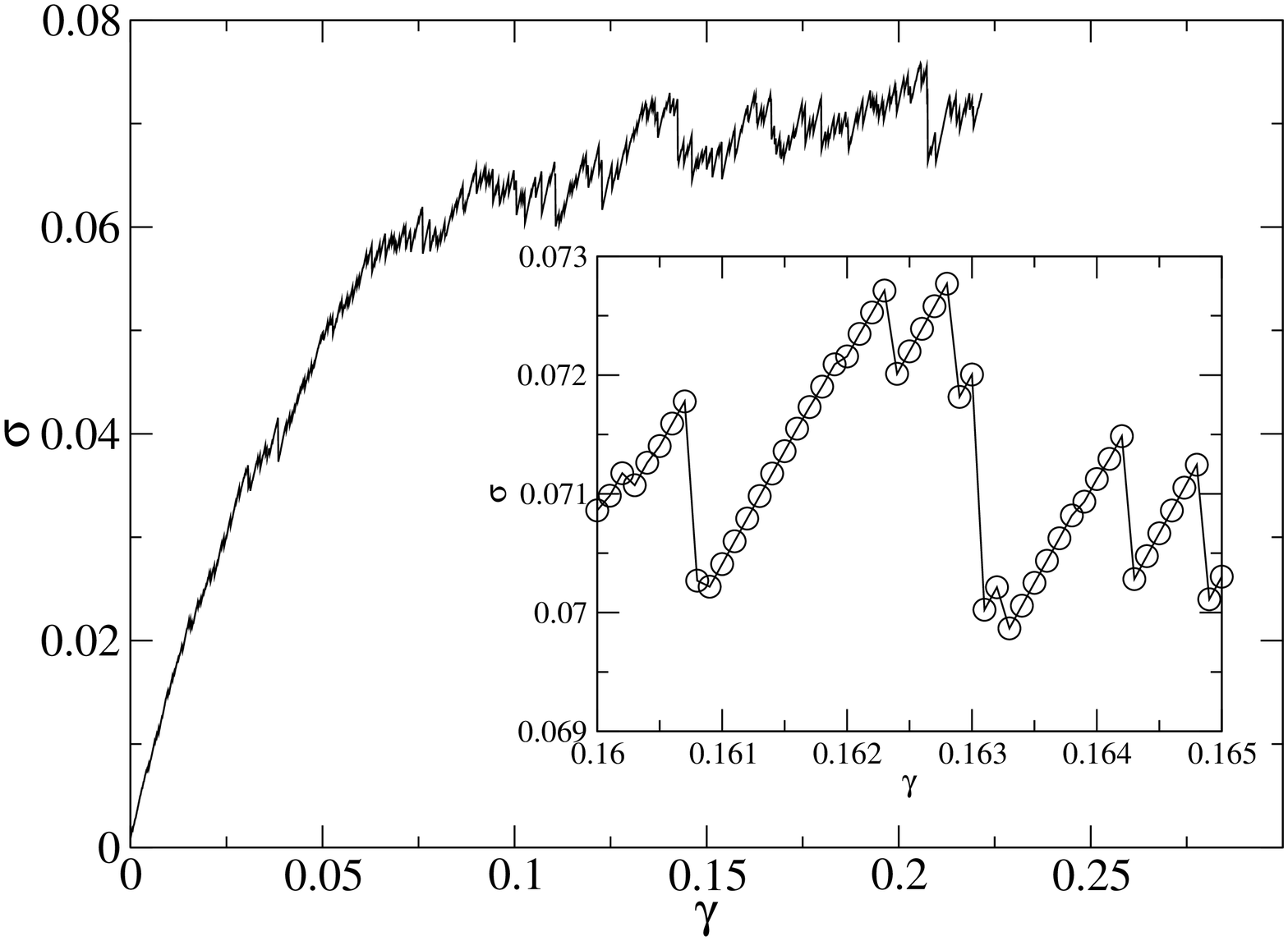}
  \caption{Stress vs. strain curve for a 200x200 system of harmonic discs.  The event at $\gamma=.1631$ will be discussed further below in section~\ref{sec:plasticEvent}.  Note the smooth, roughly linear elastic segments interrupted by the discrete plastic events.}
  \label{fig:stressStrainLong}
\end{figure}

From the preceeding picture, we understand that as a small strain rate is applied to such an athermal system, its response involves two types of behavior. Usually, the system smoothly and reversibly follows the continuous shear-induced changes of a single minimum as described above. Occasionaly, the occupied minimum vanishes, and the system has to relax toward an entirely new minimum in configuration space. This intermittent behavior shows up clearly when energy or stress is plotted against strain as in figure~\ref{fig:stressStrainLong}. On this figure, smooth segments correspond to reversible, elastic, changes of a particular energy minimum in configuration space; they are interrupted by discontinuous jumps, which corresponding to the shear induced annihilation of that minimum with a barrier. It is only during these jumps that energy is dissipated and across the jumps that irreversiblity may enter.

The energy dissipation which occurs during these discrete events will take some finite time, $\tau_{\rm dissip.}$, and in order for the system to track the changes in the potential energy landscape, it must be driven at a slow enough rate such that these discrete events have sufficient time to complete: $\dot\gamma<<1/\tau_{\rm dissip.}$.
Although the mechanisms of energy dissipation are system specific, it is reasonable to expect that material response in the quasi-static limit is largely determined by the existence of a potential energy landscape and not by the detailled mechanisms of energy dissipation.~\cite{Radjai2002,daCruz2005,Tewari1999,Lacks2002}. We will see, however, that in the AQS limit, the energy dissipation is strongly intermittent, and plastic jumps seen on figure~\ref{fig:stressStrainLong} exhibit a broad range of amplitudes. As we will see, the typical amount of dissipation in an event, and accordingly the timescales which would be associated with that energy dissipation, show strong finite size effects. We should thus keep in mind that since $\tau_{\rm dissip.}$ might depend on the system size, it is probably only justified to speak of the quasi-static limit as a formal $\dot\gamma\to0$ limit, for a fixed finite system size.

We see that the AQS limit entails three limits: zero temperature, zero strain limit, and the large size, thermodynamic limit. From the preceeding discussion it appears that the AQS limit holds when these limits are taken in the order: $T\to0$, then $\dot\gamma\to0$, then $L\to\infty$~\footnote{Physically, the order of limits should remind us that  for fixed low temperature ordinary thermal relaxation may start to compete with the effects of the shear at vanishing strain rates~\cite{Yamamoto1998,Angelani2002}, with the system thus leaving the athermal regime on approach to $\dot{\gamma}=0$ at finite $T$.}. 


As long as the system remains in a convex region surrounding a minimum of the potential energy landscape, as it does along the continuous elastic branches, the precise form of the energy minimization method should have no impact and the system will return to the local energy minimum after it is perturbed by the externally imposed deformation.
On the other hand, when the system is driven past a limit of stability, as in the third frame of figure~\ref{fig:landscapeCartoon}, the precise method of energy minimization could, in principle, have an impact on the selection of a new minimum in which to reside as the system "rolls downhill" away from the minimum which was just destroyed.

The physical mechanisms for energy dissipation are modeled differently for different systems.
In Durian's bubble model~\cite{Durian1995,Durian1997}, bubbles exert drag forces on each other proportional to their relative velocities; in Cundall and Strack's model for granular materials~\cite{Cundall1979}, grains dissipate energy via a viscous dashpot connected in parallel with the springs which repel the particles; in simulations of molecular or metallic glasses~\cite{EvansMorrissBook}, one generally uses some sort of fictitious viscous thermostat to control the temperature in the system.
Historically the AQS procedure has been implemented using some efficient energy minimization scheme, such as the non-linear conjugate gradient method, and we proceed along these lines.
However, one may hope that the details of the minimization technique, in particular, the nature of the viscosity and the existence of finite inertia, do not change the general picture that can be drawn from AQS simulations; a point of view we tentatively adopt here.


This point of view finds some support in comparative studies of MD and AQS simulations,~\cite{Lacks2002}. Lacks has shown that the effective viscosity and diffusivity of the MD simulations extrapolates to the AQS results in the zero temperature limit for a low strain rate. In related work, Yamamoto and Onuki~\cite{Yamamoto1997,Yamamoto1998} have shown that the viscosity and diffusivity in similar simulations can be understood in terms of spatially heterogeneous dynamics which themselves are controlled by a critical point at $T=0,\dot{\gamma}=0$. 
As we will see, these observation seem to be in agreement with ours and provide further indication that AQS simulation are a valid limit of MD simulations. To the best of our knowledge, however, no such explicit connection between MD and AQS has been shown for foam or granular models, but we consider it likely that analogous results would be obtained in these models in the limit of vanishing strain rate.
We will therefore assume that the AQS procedure is equally applicable to the wet foams and frictionless granular systems, in addition to the metallic glasses for which it has traditionally been considered to apply.


\subsection{Numerical Details}

In this work we deal exclusively with 2D systems.
Bi-disperse mixtures are used to inhibit crystallization.
The mixtures used throughout have particles with radii: $r_{L}=.5$ and $r_{S}=.3$~\footnote{
We have incorrectly stated the ratio of radii in our previous works~\cite{Maloney2004,Maloney2004a,Lemaitre2005}.}
and a number ratio of: $N_{L}=N_{S}\frac{1+\sqrt{5}}{4}$.
The mixture is similar to that used by Falk and Langer and was found sufficient to inhibit crystallization.
The systems are prepared via a zero temperature quench from an initially random state as in references~\cite{O'Hern2002,O'Hern2003}.  
We suppose that the systems lose memory of their initial preparation before a strain of unity and use the strain of unity as a starting point for tabulating statistical properties of the steady flow.
No segregation is detectable within the 200\% window over which the systems are strained, but it would be difficult to rule out effects which occur on very long strain scales.
Lees-Edwards boundary conditions are used throughout~\cite{EvansMorrissBook}.
The athermal quasistatic dynamics algorithm described above is implemented using strain steps of size $10^{-4}$ which, as will be discussed below, is small enough such that all loading curves have well resolved elastic segments even for the largest samples simulated. 

Two different minimization algorithms from the conjugate gradient family were employed~\cite{NocedalWrightBook}: the non-linear Pollak-Riberi conjugate gradient minimizer and the truncated Newton linearized conjugate gradient minimizer.   
For the non-linear Pollak-Riberi conjugate gradient minimizer, we used the routine as implemented in the GNU Science Library~\cite{gsl}.
For the truncated Newton linearized conjugate gradient minimizer, we used an Armijo backtracking algorithm~\cite{NocedalWrightBook} with a sufficient decrease parameter of $.1$ and the linear conjugate gradient routine as implemented in the Iterative Template Library~\cite{itl} adapted by us to perform truncation~\cite{NocedalWrightBook} upon encountering curvatures less than $10^{-12}$.
We found the latter procedure to be more robust and efficient.
The simulations of the single 50x50 system of Lennard-Jones particles discussed in section~\ref{sec:elasticSegments} and the 200x200 system of harmonically interacting particles discussed in section~\ref{sec:plasticEvent} utilized the former minimization algorithm, while the data runs which were used for the analysis in section~\ref{sec:scaling} utilized the latter.
Both algorithms gave statistically identical results when run on an ensemble of 50x50 harmonic systems.

The potential energy functions were pairwise additive central force laws. 
Three different force laws were employed: a standard 6-12 Lennard-Jones interaction, a harmonic, repulsive spring force, and a non-linear hertzian repulsive spring force~\footnote{Note that, in hertzian contact theory, discs in 2D should interact via a harmonic spring.  The hertzian spring in our 2D model is not meant to be a quantitatively accurate representation for any particular 2D granular material, but rather is used to show the independence of the emergent behavior on generic non-linearities in the interaction.}.
The Lennard-Jones energy was truncated at a distance of 2 particle diameters, and linear and quadratic terms were added to ensure continuity up through second derivatives at the cutoff to avoid pathologies in the minimization routine.

The pair interactions read:
\begin{eqnarray*}
  U_{\text{Harm}}(r_{ij})&=&(1-s_{ij})^2\theta(1-s_{ij})\\
  U_{\text{Hertz}}(r_{ij})&=&(1-s_{ij})^{5/2}\theta(1-s_{ij})\\
  U_{\text{LJ}}(r_{ij})&=&\left(s_{ij}^{-12}-2s_{ij}^{-6}+As_{ij}+Bs_{ij}^2\right)\theta(2-s_{ij})
\end{eqnarray*}
where $\theta$ is the unit step, and $A$ and $B$ are the coefficients which force continuity of the first and second derivatives at the cutoff in the Lennard-Jones potential.
$s_{ij}$ is the dimensionless separation between particles $i$ and $j$,
\[s_{ij}=\frac{r_{ij}}{R_i+R_j}\]
where $R_i$ is the radius of particle $i$.
\section{Elastic Breakdown and Plastic Nucleation}
\label{sec:elasticSegments}
As described above, trajectories in AQS are composed of smooth, reversible elastic segments which are separated by discontinuous, irreversible plastic events.
The smoothness of the elastic segments allows us to obtain analytical results regarding the singularity at the onset of a plastic event and to analyze the real-space structure of the critical mode of the system which is responsible for nucleating the subsequent plastic event. 

\subsection{Analytical Framework}
We first briefly review the formalism developed in~\cite{Lemaitre2005} and recall the key results.
The basic principle underlying this framework is that, owing to the smoothness of the energy during the elastic segments, exact analytical expressions for the trajectory can be written by requiring that the system track the local energy minimum as its location in configuration space changes due to the imposed shear strain.

The resulting equations governing the trajectory of the particles were found to be\footnote{We use the convention that Latin indices indicate Cartesian directions, while Greek indices label particles.}
:
\begin{equation}
  \mathring{v}_{i\alpha}\doteq\frac{d\mathring{r}_{i\alpha}}{d\gamma}=-H^{-1}_{i\alpha j\beta}\Xi_{j\beta}
  \label{eq:QSD}
\end{equation}
The open circle over the $r$ in equation~(\ref{eq:QSD}) is to indicate that the derivative is to be taken in the frame which is co-moving with the simple shear;
that is:
\begin{eqnarray}
\mathring{x}(\gamma) & = & x(\gamma)\\
\mathring{y}(\gamma) & = & y(\gamma)-\gamma x(\gamma=0).
\end{eqnarray}
Thus, equation~(\ref{eq:QSD}) describes the \emph{non-affine} component of the motion, and the full motion of the particles in the laboratory frame is given by the imposed homogeneous shear plus the correction term embodied in equation~(\ref{eq:QSD}).
$H_{i\alpha j\beta}$ is the so-called ``hessian matrix'' or ``dynamical matrix'' -- the second derivatives of the energy: $H_{i\alpha j\beta}\doteq\frac{\partial^{2}U}{\partial r_{i\alpha}\partial r_{j\beta}}$.
$\Xi_{i\alpha}$ is the derivative of the net force on particle $i$ with respect to strain --- or, equivalently, the derivative of the stress contribution of particle $i$ with respect to a change in its position: $\Xi_{i\alpha}\doteq\frac{\partial^{2}U}{\partial \gamma \partial r_{i\alpha}}$. 
As discussed in detail in~\cite{Lemaitre2005}, $\Xi_{i\alpha}$ vanishes for configurations with local symmetry, and, as such, is a measure of the local configurational disorder about particle $i$.

The stress is defined as the total derivative of the energy with respect to $\gamma$ while enforcing mechanical equilibrium of the particles via equation~(\ref{eq:QSD}).
\begin{equation}
  \sigma\doteq\frac{dU}{d\gamma}=\frac{\partial U}{\partial \mathring{r}_{i\alpha}} \frac{d\mathring{r}_{i\alpha}}{d\gamma} + \frac{\partial U}{\partial \gamma}=\frac{\partial U}{\partial \gamma}
  \label{eq:sigma}
\end{equation} 
where the final equality holds because of mechanical equilibrium.
So we see that the fact that the particles do not follow affine trajectories does \emph{not} give rise to any corrections to the stress.
This is not true, however, for the shear modulus.

To find the shear modulus, we simply differentiate yet one more time, again with a total derivative which should be understood to be taken while enforcing the corrections given in equation~(\ref{eq:QSD}).
\begin{equation}
  \mu\doteq\frac{d\sigma}{d\gamma}=\frac{\partial^{2}U}{\partial \gamma ^{2}} - \Xi_{i\alpha}H^{-1}_{i\alpha j\beta}\Xi_{j \beta}= \mu_{a}-\mu_{na}
  \label{eq:mu}
\end{equation}
In equation~(\ref{eq:mu}), $\mu_{a}\doteq\frac{\partial^{2}U}{\partial \gamma ^{2}}$ is the term which arises from the Born expression for pure affine deformation~\cite{BornHuangBook}, and $\mu_{na}\doteq\Xi_{i\alpha}H^{-1}_{i\alpha j\beta}\Xi_{j \beta}$ is the term which comes from considering the non-affine corrections.
Note that the corrected modulus is always less than the naive Born expression.
The microscopic analytical expressions for $\sigma$, $\mu_{a}$, $H_{i\alpha j\beta}$ and $\Xi_{i\alpha}$ in the case of pairwise interacting systems can be found in reference~\cite{Lemaitre2005}, but we emphasize that the equations~(\ref{eq:QSD}),~(\ref{eq:sigma}), and~(\ref{eq:mu}) are completely general and valid for any arbitrary n-body interaction potential (e.g. embedded atom methods~\cite{Daw1983,Daw1984}, or potentials for silicon~\cite{Stillinger1985}).

So what is the structure in real space of the response during the continuous elastic segments?
To illustrate, we now focus on a typical elastic segment in one particular 50x50 Lennard-Jones sample. 
For any particular configuration, we can efficiently solve equation~(\ref{eq:QSD}) via some iterative method.
We use the conjugate gradient algorithm exactly as implemented in the Iterative Template Library~\cite{itl} with a relative tolerance of $10^{-8}$.
This method, utilizing the analytical form for the elastic response, should be preferred over the alternate method of explicitly shearing the system by a small, finite amount then reminimizing the energy.
The latter method essentially amounts to using a finite difference in lieu of a derivative whose analytical form we know.
In reference~\cite{Tanguy2002}, it was found using the alternate method of explicit shear that when taking a small enough strain step to remain in the linear regime, the energy had to be computed to quadruple precision.
No such special measures should be necessary in directly solving equation~(\ref{eq:QSD}).

\begin{figure}
\resizebox{!}{.4\textwidth}{{\includegraphics{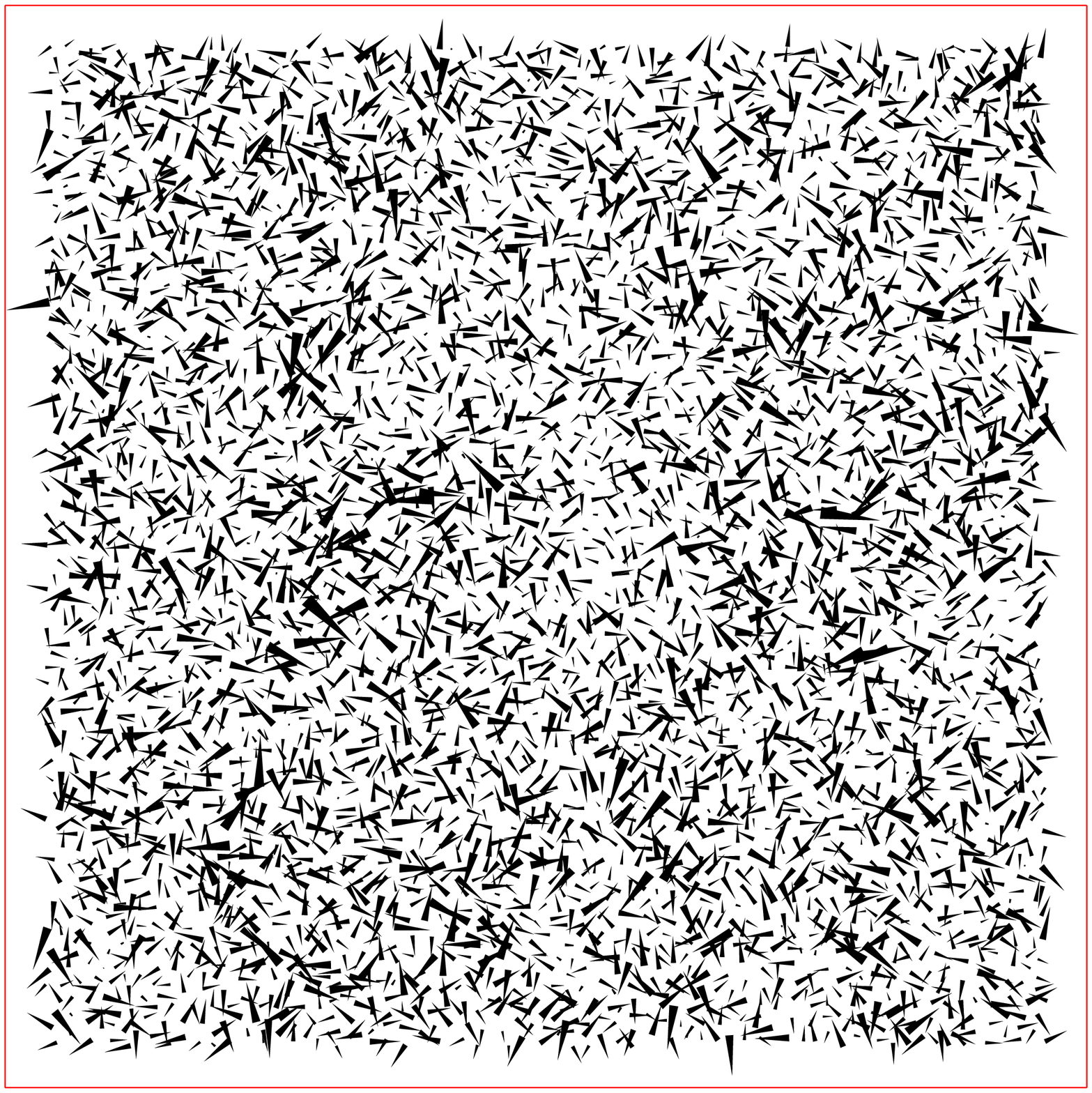}}}\\
\resizebox{!}{.4\textwidth}{{\includegraphics{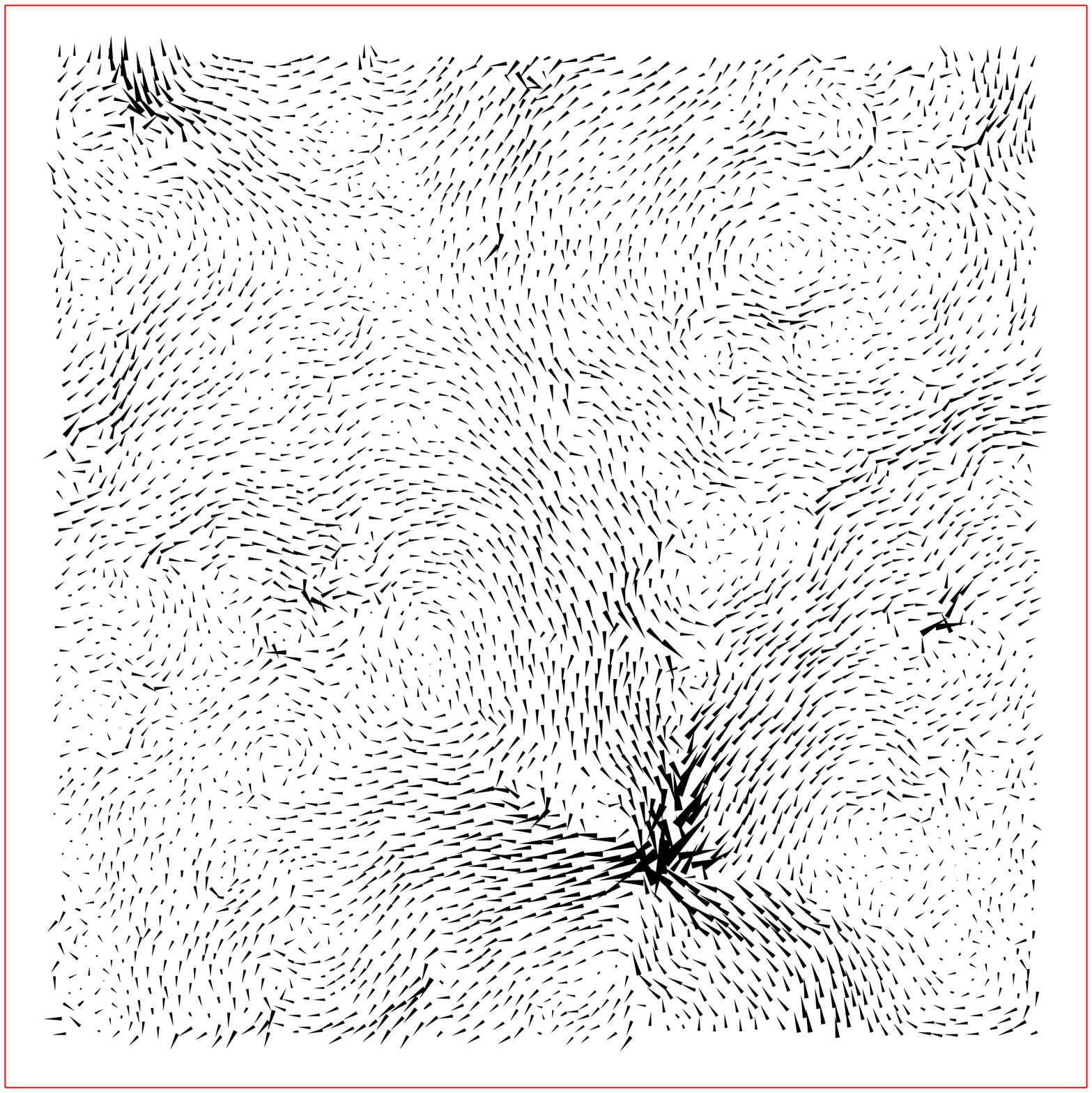}}}
\caption{
The particular force in response to homogeneous shear,
$\vec{\Xi}$ (top),
and the non-affine velocity (or ``displacement'') field,
$\mathring{v}_{i\alpha}$ (bottom),
at a strain configuration,
$\gamma=0.2945$,
or $\gamma_c-\gamma\sim 10^{-4}$.
}
\label{fig:xiTyp}
\end{figure}

Figure~\ref{fig:xiTyp} shows the fields $\Xi_{i\alpha}$ and $\mathring{v}_{i\alpha}$ at a value of the strain which is roughly a distance of $10^{-4}$ from the next catastrophic event.
Note that the $\Xi$ field appears essentially random, as expected, based on its role as a measure of local configurational disorder.
$\mathring{v}_{i\alpha}$, on the other hand, should depend strongly on the low modes in the spectrum of $H$, as can be seen directly from equation~(\ref{eq:QSD}).
It exhibits striking correlations in both compression and shear.

The strongly spatially correlated behavior apparent in this elastic response field (and accordingly $|\mu_{na}|$) should be contrasted with the lack of any correlation beyond a length of a few particle diameters in any of the other mechanical quantities such as $\mu_{a}$, energy, pressure, shear stress, or vonMises stress.  
The importance of such non-affine corrections to elastic behavior has been realized in the context of experiments and simulations on emulsions and foams reported by Liu and co-workers~\cite{Liu1996} and Langer and Liu~\cite{Langer1997}.
More recently, Wittmer, Tanguy, and co-workers \emph{et. al.} have conducted a comprehensive study of the contribution to elasticity of the non-affine rearrangements in a Lennard-Jones system~\cite{Wittmer2002,Tanguy2002}.
However, in this work, we will be more interested in the role of the non-affine response on approach to the end-points of the elastic segments and its role in nucleating the plastic cascades rather than its role in renormalizing the elastic moduli away from the plastic events.

\subsection{Approaching Catastrophes}
Equation~(\ref{eq:QSD}), which describes the elastic segments, breaks down precisely when the local minimum vanishes; \emph{i.e.} when the potential energy surface develops a direction of zero curvature, along which the minimum collides with a first order saddle as in the cartoon of figure~\ref{fig:landscapeCartoon}.
This scenario, with a \emph{single} control parameter destabilizing a \emph{single} degree of freedom, is the simplest possible type of bifurcation --- known as a \emph{fold} in bifurcation and catastrophe theory or a \emph{tangent bifurcation} in dynamical systems theory~\cite{ArnoldBook}---,
and we will see how this breakdown dictates the scalings of various quantities with strain at the onset of the plastic events.

\begin{figure}
\includegraphics[angle=-90,width=.45\textwidth]{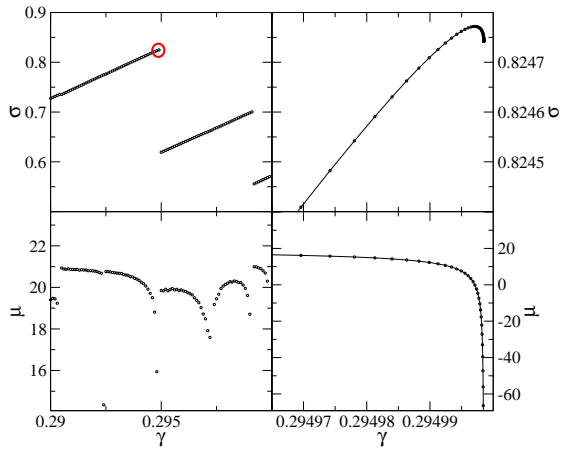}
\caption{\label{fig:endpointBlowup}
Stress (top) and shear modulus (bottom) for a small strain interval.
Left: fixed strain steps of size $10^{-4}$ using the standard energy minimization algorithm;
Right: convergence to the yield point using a decreasing strain step size and the modified linesearch algorithm described in the appendix.}
\end{figure}

Since $\Xi$ depends only on local information about the near-neighbor particle configurations, we expect that it may be treated as roughly constant in the neighborhood of one of these catastrophic events, whence, according to equation~(\ref{eq:QSD}) the elastic response field will start to diverge along the direction of the zero curvature.
As the low curvature direction gets flatter and flatter, eventually the non-affine correction to the shear modulus dominates the Born term in equation~(\ref{eq:mu}) and the net modulus becomes negative.
At this point the stress starts to decrease as a function of strain, and the system is unstable against any applied stress. 
These configurations are accessible to us because strain -- and not stress -- is controlled.
Eventually, the curvature will go to zero, and $\mathring{v}$ and $\mu_{na}$ will diverge.

We now proceed to isolate the singular behavior at the catastrophic points.
In the lowest order non-trivial Taylor expansion for the energy at the transition point, we must include a higher order term for the critical direction, as the quadratic term vanishes.
Generically, we expect a cubic term to remain~\cite{ArnoldBook}:
\begin{equation}
  U\sim As^{3}_{0}+\frac{1}{2}s_{i\alpha}\mathcal{H}_{i\alpha j\beta}s_{j\beta}+\delta\gamma(\Sigma+\Xi_{i\alpha}s_{i\alpha})+\frac{\mu_{a}}{2}\delta\gamma^{2}
  \label{eq:expansion}
\end{equation} 
$A,\Sigma,\mathcal{H}_{i\alpha j\beta},\Xi_{i\alpha}$, and $\mu_{a}$ are constants which are evaluated \emph{at} the transition.
The physical interpretation of all but $A$ are discussed at length in reference~\cite{Lemaitre2005}.
$s_{i\alpha}$ denotes the displacement of the reference coordinate away from the critical configuration: $s_{i\alpha}\doteq\mathring{r}_{i\alpha}(\gamma)-\mathring{r}_{i\alpha}(\gamma_{c})$.
$s_{0}$ denotes the projection onto the critical mode, $s_{0}\doteq s_{i\alpha}\Psi^{0}_{i\alpha}$, where $\Psi^{0}_{i\alpha}$ is the unit vector which lies in the (non-translational) null space of $\mathcal{H}_{i\alpha j\beta}$.
So we have:
\begin{equation}
  -F_{i\alpha}\doteq\frac{\partial U}{\partial s_{i\alpha}}=3As^{2}_{0}\Psi^{0}_{i\alpha}+\mathcal{H}_{i\alpha j\beta}s_{j\beta}+\delta\gamma \Xi_{i\alpha}
\label{eq:force}
\end{equation}
\begin{equation}
  H_{i\alpha j\beta}=\frac{\partial^{2}U}{\partial s_{i\alpha}\partial s_{j\beta}}=\mathcal{H}_{i\alpha j\beta}+6As_{0}\Psi^{0}_{i\alpha}\Psi^{0}_{j\beta}
\end{equation}
Requiring equation~(\ref{eq:force}) to be zero (which is equivalent to applying equation~(\ref{eq:QSD})), and since $\Psi^{0}_{i\alpha}$ lies in the null space of $\mathcal{H}_{i\alpha j\beta}$, we have:
\begin{equation}
  \mathcal{H}_{i\alpha j\beta}s_{j\beta}=-\delta\gamma \Xi_{i\alpha}
\label{eq:forceEqZero1}
\end{equation}
and
\begin{equation}
  3As^{2}_{0}=-\delta\gamma \Xi_{0}
\label{eq:forceEqZero2}
\end{equation}
We may solve equation~(\ref{eq:forceEqZero1}) for $s_{i\alpha}$ up to the $d$ uniform translational modes and the critical mode.
Equation~(\ref{eq:forceEqZero2}) then provides the solution for the critical mode: $s_{0}=\sqrt{-\delta\gamma \Xi_{0}/3A}$.
The motion along this critical mode experiences a square root singularity, characteristic of the simple fold catastrophe, while the motion along the higher modes is smooth at the level of the expansion~(\ref{eq:expansion}).

This singular behavior for $s_{0}$ induces singular behavior in the modulus and stress.  Expanding equation~(\ref{eq:sigma}), we have:
\begin{equation}
  \sigma=\Sigma-\delta\gamma(\tilde{\Xi}_{i\alpha}\mathcal{H}^{-1}_{i\alpha j\beta}\tilde{\Xi}_{j\beta})-\sqrt{\frac{-\delta\gamma \Xi^{3}_{0}}{3A}}+\mu_{a}\delta\gamma
  \label{eq:muSingular}
\end{equation}
where $\tilde{\Xi}_{i\alpha}$ is $\Xi_{i\alpha}$ with the critical component projected out: $\tilde{\Xi}_{i\alpha}=\Xi_{i\alpha}-\Psi^{0}_{i\alpha}\Psi^{0}_{j\beta}\Xi_{j\beta}$. 
Expanding equation~(\ref{eq:mu}), or, equivalently, taking the $\gamma$ derivative of equation~(\ref{eq:muSingular}), we get:
\begin{equation}
  \mu\doteq\frac{d\sigma}{d\gamma}=\mu_{a}-\tilde{\mu}_{na}-(-\delta\gamma)^{-1/2}\sqrt{\frac{\Xi^{3}_{0}}{12A}}
\end{equation}
where $\mu_{a}$ is the Born contribution to the modulus as in equation~(\ref{eq:mu}), $\tilde{\mu}_{na}=\tilde{\Xi}_{i\alpha}\mathcal{H}^{-1}_{i\alpha j\beta}\tilde{\Xi}_{j\beta}$ is the non-singular part of the non-affine correction, and the remainder is the isolated singular piece. 

To check these predictions, we perform a careful convergence to a particular catastrophic point in the system shown above in figure~\ref{fig:xiTyp}.
During the linesearch portion of the minimization routine, we converged first to a minimum of force and subsequently to a minimum of energy along the line and found this procedure to produce robust convergence to the transition point.

\begin{figure}
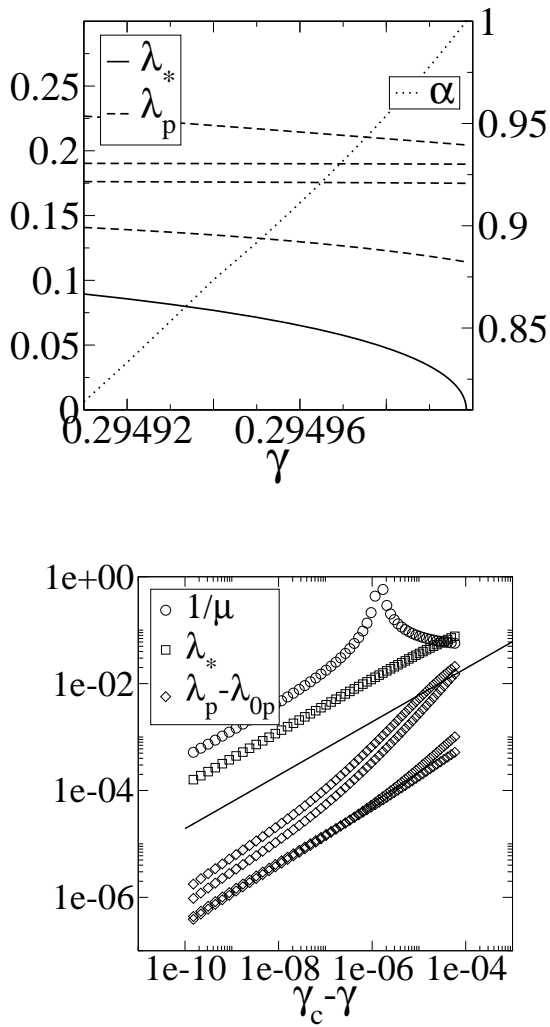

\includegraphics[width=.4\textwidth]{endpointEigenvals.eps}

\vspace{1cm}
\includegraphics[width=.35\textwidth]{endpointAllLog.eps}
\caption{
All data shown here for the same trajectories plotted above in figure~\ref{fig:endpointBlowup}b.
a) relative participation of the lowest normal mode in the non-affine elastic displacement field,
$\alpha^*\doteq (\Psi^{0}_{i\alpha}\mathring{v}_{i\alpha})^{2}/(\mathring{v}_{i\alpha}\mathring{v}_{i\alpha})$
(dotted); lowest eigenvalue of the dynamical matrix (solid); next several eigenvalues (dashed).  b) In log-log scale
(as a guide to the eye, the thick black line is $\sqrt{\gamma_c-\gamma}$):
$1/\mu$ (circles); lowest eigenvalue (squares); next several eigenvalues minus their terminal values (diamonds).
}
\label{fig:convergence}
\end{figure}

Figure~\ref{fig:endpointBlowup} shows the stress and modulus for the same system as in figure~\ref{fig:xiTyp} upon approach to the singularity.
Zooming in on the endpoint of the elastic segment, we see that the qualitative predictions of the theoretical arguments are borne out; namely that the stress reaches a maximum precisely as the modulus becomes negative.
We further note that the Born contribution to the modulus is essentially constant on this region of $\gamma$ (not shown).

In figure~\ref{fig:convergence}a, we plot the several smallest eigenvalues and the participation of the lowest mode, $\alpha(\gamma)$.
In agreement with Malandro and Lacks, we find that a \emph{single} eigenvalue vanishes.
In the window of strain shown, the participation of the lowest mode goes linearly from about $.85$ at a distance of about $\delta\gamma\sim 10^{-4}$ to nearly unity at a distance of $\delta\gamma\sim 10^{-10}$.

Figure~\ref{fig:convergence}b shows the stress, the inverse of the modulus, and several higher eigenvalues as functions of strain on a log-log scale.
The critical strain is determined to within $10^{-12}$ (not shown), and this terminal value is used to measure $\delta\gamma\doteq\gamma_{c}-\gamma$ for configurations down to $10^{-10}$ (shown).
All quantities exhibit the same $\sqrt{\delta\gamma}$ behavior at small $\delta\gamma$. 
This was predicted above for the modulus and critical curvature, however, the higher curvatures are constants at the order of the expansion~(\ref{eq:expansion}). 
We can rationalize the $\sqrt{\delta\gamma}$ behavior for the higher modes by considering that the total derivative of \emph{any} function, $f(s_{i\alpha})$, should be dominated by the singular behavior of $s_{0}$: 
\[\frac{df}{d\gamma}\sim\frac{\partial f}{\partial s_{0}}\frac{ds_{0}}{d\gamma}\sim\frac{\partial f}{\partial s_{0}}(-\delta\gamma)^{-1/2}\]  

\subsection{The Critical Mode}
The real space structure of a localized plastic event is a key input into coarse-grained models of plasticity~\cite{Bulatov1994,Bulatov1994a,Bulatov1994b,Langer2001,Baret2002,Onuki2003,Picard2005} and different supposed forms could lead to different emergent behavior in these models.
A real space analysis of the incipient failure mode, although it does not correspond to a \emph{complete} shear transformation, but rather to the \emph{onset} of one, gives some insight into the form of an elementary plastic event.

\begin{figure}
  \includegraphics[width=.4\textwidth]{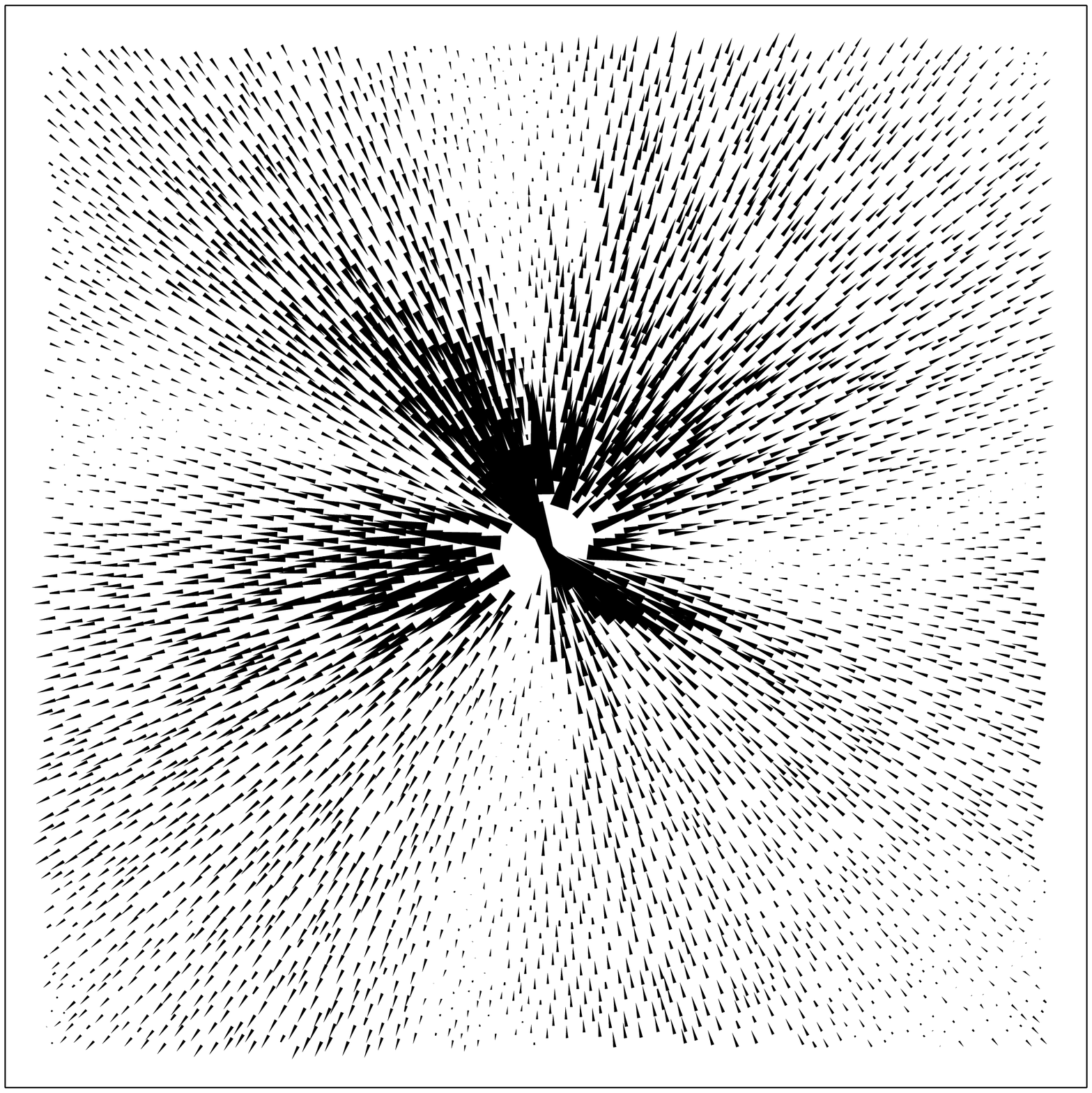}
  \includegraphics[width=.4\textwidth]{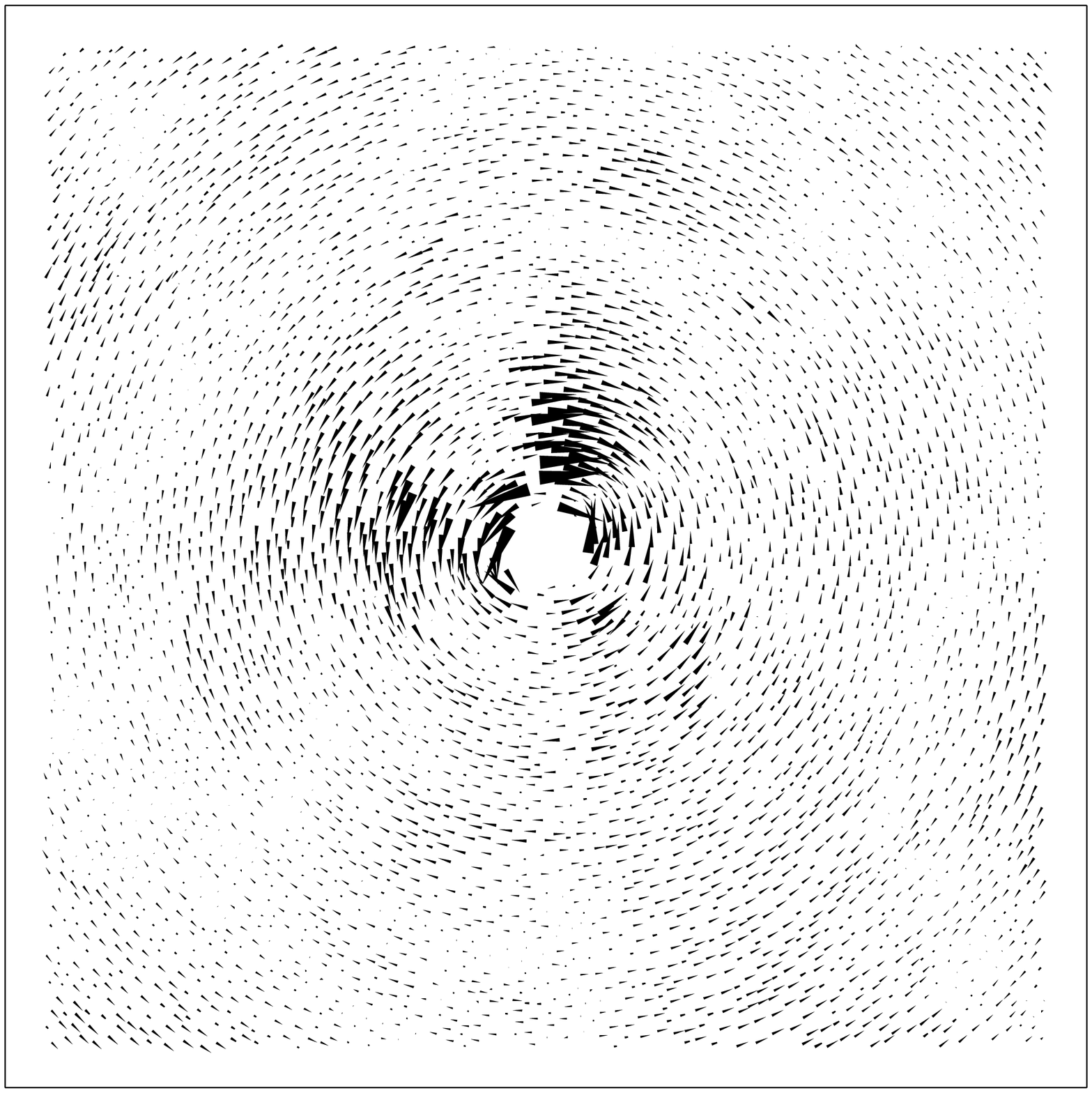}
  \caption{The radial and tangential projections of the non-affine displacement field, $\mathring{v}_{i\alpha}$, at $\gamma_c - \gamma \sim 10^{-10}$ transformed back to the rectilinear frame and centered on the vector with largest magnitude as described in the text.
The center corresponds to the quadrupolar pattern which is starting to develop in the lower right of figure~\ref{fig:xiTyp} at a strain of $\gamma_c-\gamma\sim 10^{-4}$.
A core region of radius 2 is excluded from the image (and subsequent analysis).
Scale is arbitrary: the field is normalized to unity.
At this strain, $\mathring{v}_{i\alpha}$ is essentially indistinguishable from the critical eigenmode, $\Psi^{0}_{i\alpha}$.
}
  \label{fig:projections}
\end{figure}

In figure~\ref{fig:projections}, we show the field $\mathring{v}_{i\alpha}$ at $\delta\gamma\sim 10^{-10}$, at which point it is almost entirely aligned with the critical mode.
The geometry is predominantly quadrupolar; a core region with outward particle velocities along the tensile axis and inward particle velocities along the compressive axis.
We stress that the eigenmodes themselves have changed little in the window of strain from $\delta\gamma\sim 10^{-4}$ to $\delta\gamma\sim 10^{-10}$ as we have approached the incipient failure event, and the change in $\mathring{v}_{i\alpha}$ comes almost entirely from the change in the relative weightings of the modes, with the quadrupole in the lower right of figure~\ref{fig:xiTyp}b becoming dominant at the critical strain.

In constructing figure~\ref{fig:projections} we have performed two transformations.
First, we have applied an inverse affine transformation to the locations of the particles and their displacement vectors, $\mathring{v}_{i\alpha}$:
\[\begin{array}{l}
r_{y}\rightarrow r_{y}\\ 
r_{x}\rightarrow r_{x}-\gamma r_{y}\\
\mathring{v}_{y}\rightarrow \mathring{v}_{y}\\
\mathring{v}_{x}\rightarrow \mathring{v}_{x}-\gamma \mathring{v}_{y}
\end{array}\]
In this frame, the system maps onto itself under a purely vertical or purely horizontal shift by the box length, a property one which would be lost in a Lees-Edwards cell at finite strain. 
Further, we have moved to a co-ordinate system, $(r,\theta)$, which is centered on the particle with the largest vector; for simplicity we have taken the center of the core to be on the particle with largest $\mathring{v}$.
Also note that we have considered separately the radial and tangential projections and excluded a core region of radius 2.

It is natural to ask whether the eigenmode is analogous to the displacement field generated in a linear, isotropic, elastic medium by some disturbance at the core.
To map the discrete vector field onto a continuous field, we divide the system into annuli of width 2 starting at a radius of 2 outside of the nominal core.
In each annulus, we project onto circular harmonics by summing:
\[\begin{array}{l}
\tilde{v}_{r}(r;n)=\displaystyle\sum_{j\in\mathcal{A}(r)}v_{rj}e^{in\theta_{j}}\\
\\
\tilde{v}_{\theta}(r;n)=\displaystyle\sum_{j\in\mathcal{A}(r)}v_{\theta j}e^{in\theta_{j}}
\end{array}
\]
(where $j$ indexes the particles in a given annulus ) and normalizing each term by the number of particles in the annulus.

It is found through this decomposition and is obvious from simple visual inspection of the field, that the $n=2$ (quadrupole) contribution dominates and has a phase angle which roughly gives extension along $y=x$ and compression along $y=-x$ (there is a slight clockwise departure which can be seen in figure~\ref{fig:projections})
while the tangential component is roughly aligned at $y=0,x=0$.

If the material behaves as a linear homogeneous isotropic elastic solid outside of some core region, we would expect, recalling the form of the quadrupolar space of solutions of the 2D Navier-Lam\'{e} equation, that~\cite{SlaughterBook}:
\begin{eqnarray}
v_{r}(r;2)=\frac{2A}{r^3}+\frac{(1+\kappa)B}{r} \label{eq:radialQuad}\\
v_{\theta}(r;2)=\frac{2A}{r^3}+\frac{(1-\kappa)B}{r} \label{eq:azimuthQuad}
\end{eqnarray}
where $\kappa$ is the ratio of Lam\'e constants: $\kappa=\frac{\lambda}{2(\lambda+\mu)}$
and the $\theta$ dependence of the radial and azimuthal fields should be understood to have a relative phase of 45 degrees.
\begin{figure}
  \includegraphics[width=.45\textwidth]{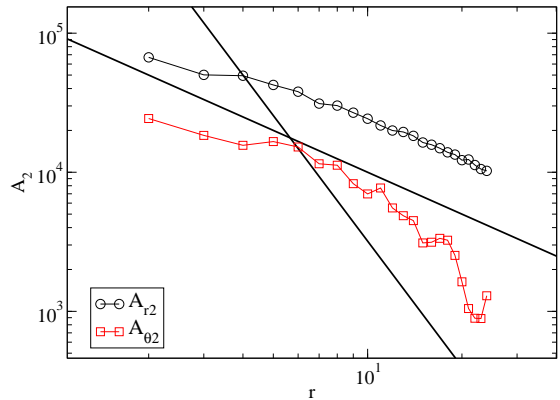}
  \caption{The magnitudes of the quadrupolar ($n=2$) projections of the radial, $A_{r2}$ (black circles), and azimuthal $A_{\theta 2}$ (red squares), components of of the critical mode.  Solid lines are $r^{-1}$ and $r^{-3}$.}
  \label{fig:critModeLogLog}
\end{figure}
In figure~\ref{fig:critModeLogLog}, we plot the magnitude of the quadrupolar sector for $v_r(r)$ and $v_\theta(r)$.

The radial field seems to be consistent with a pure $r^{-1}$ behavior, while the azimuthal field becomes too noisy to determine whether it follows any particular power law.
For a general quadrupole, one would expect a crossover to $r^{-3}$ behavior at small $r$, and it is evident that this crossover length is small if the $r^{-3}$ term is even present at all.
Furthermore, we may attempt to extract a value of $\kappa=(v_r-v_{\theta})/(v_r+v_{\theta})$.
For radii less than $10$, the azimuthal field is not too noisy, and the extracted $\kappa$ fluctuates between $.4$ and $.5$ (not shown), $.4$ being roughly consistent with the value of $\kappa$ averaged over the  elastic segments.
At larger radii, the extracted $\kappa$ leaves the physically allowed regime, becoming larger than $.5$, which is not surprising given the increased noise in $v_{\theta}$ at these radii.
The agreement between this catastrophic mode and an elastic quadrupole is very reasonable.

\subsection{Discussion}
There is a subtle distinction between the incipient failure mode, which we have just measured, and a complete shear transformation.
Ideally, one would like to measure changes in the particle configurations after the system has been driven past stability and then completely relaxed. 
In our atomistic system, as in the coarse-grained models~\cite{Bulatov1994,Baret2002,Picard2005}, localized plastic events rarely occur in an isolated way, and most often the incipient failure event, such as the one measured here, triggers a subsequent plastic cascade.
As we will see below, although it is possible to make \emph{some} measurements of the elementary shear transformations which occur \emph{during} the cascades, they cannot be given as precise a meaning as the measurements of the incipient failure modes such as the one measured in this section. 

Before leaving the subject of plastic onset, we should clarify the relationship of the picture put forward here to the earlier numerical studies of Srolovitz and co-workers~\cite{Srolovitz1983}.
In those athermal simulations of metallic glasses under shear, the authors reported on stress concentrations which acted as catalysts for plasticity.
These stress concentrations were shown to vanish upon unloading, and an analogy was made with the stress fields around the tips of nascent shear cracks.
Although we will, in fact, report on a related mechanism below, we have not found such stress concentrations at the onset of plastic events.
In fact, the formalism presented above (and discussed at length in~\cite{Lemaitre2005}) makes it clear that the instability is a collective property of the system and should not necessarily be discernible by looking at strictly local quantities.
The local Born moduli, and the so-called ``site-symmetry'' parameters discussed by Egami \emph{et. al.}~\cite{Egami1980}, are closely related to the field $\Xi_{i\alpha}$ (see reference~\cite{Lemaitre2005} for the details of this relationship) which was shown to be essentially random and well behaved near the transition.
Although it is possible that there may be stronger correlations between $\Xi_{i\alpha}$ and the normal modes in other systems, such as those studied by Srolovitz and co-workers, no such correlations were found in any of the three different interaction potentials studied here in 2D, and there is no fundamental need for the stress itself to be particularly large locally in order to nucleate a plastic event.
\section{Plastic Events}
\label{sec:plasticEvent}
In the previous section, we focused on the behavior of the elastic segments of the quasistatic trajectories, and in particular, the behavior on approach to their endpoints at the onset of the plastic events.
We now discuss the processes at work during these plastic events themselves as the system searches the potential energy landscape in a fully non-linear way in search of a new inherent structure after it is driven past a threshold of stability.
A single typical event in a large system will be used as a case study.

\subsection{The Cascade Mechanism}

Already in the seminal work~\cite{Argon1979a}, Argon emphasized the analogy between a local shear transformation and the nucleation of a dislocation loop.
He cautioned that the analogy should not be taken \emph{too} literally; in a crystal, the barrier for the nucleation of a pair of dislocations is large compared to the subsequent Peierls barriers, so, once nucleated, the pair is essentially free to glide apart.
In a disordered system, on the other hand, even if one could topologically identify dislocations, the concept would be of limited utility, as there are no directions of symmetry along which the pair might glide apart.


However, the elastic-like fields which are expected to result in the surroundings of a local shear transformation should alter the probabilities of observing subsequent shear transformations in neighboring regions.
These elastic-like fields are expected to have quadrupolar symmetry (i.e. they represent elementary \emph{shear}) and resemble the elastic fields associated with the nucleation of a dislocation pair or the transformation of an Eshelby inclusion~\cite{Eshelby1957}.

Kobayashi, Maeda and Takeuchi were able to observe such elastic-like displacements of the particles surrounding the core of a single shear transformation in their computer simulations~\cite{Kobayashi1980,Maeda1981}.
Bulatov and Argon, with this picture of elastically mediated interactions in mind, constructed a stochastic model of plasticity by embedding potential shear transformation sites unifmormily in a 2D lattice~\cite{Bulatov1994,Bulatov1994a,Bulatov1994b}.
Within their model, such cascades \emph{did} emerge to play a role analogous to that of dislocation glide.
Since then, several others~\cite{Langer2001,Baret2002,Onuki2003,Picard2005} have constructed coarse-grained models along similar lines with varying assumptions about the precise microscopic details of the elastic consequences of a local plastic event.

\begin{figure}
  \includegraphics[width=.15\textwidth]{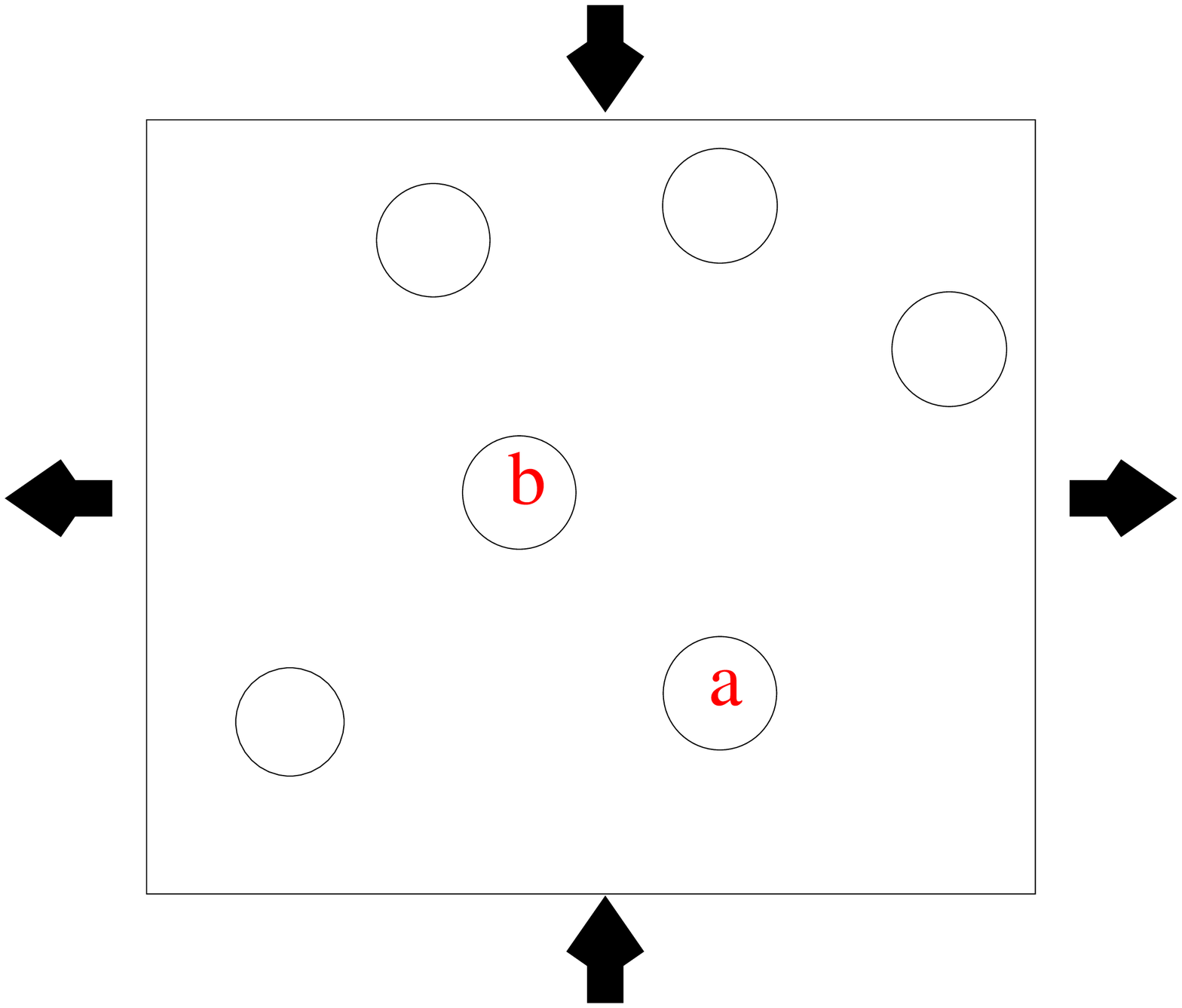}
  \includegraphics[width=.15\textwidth]{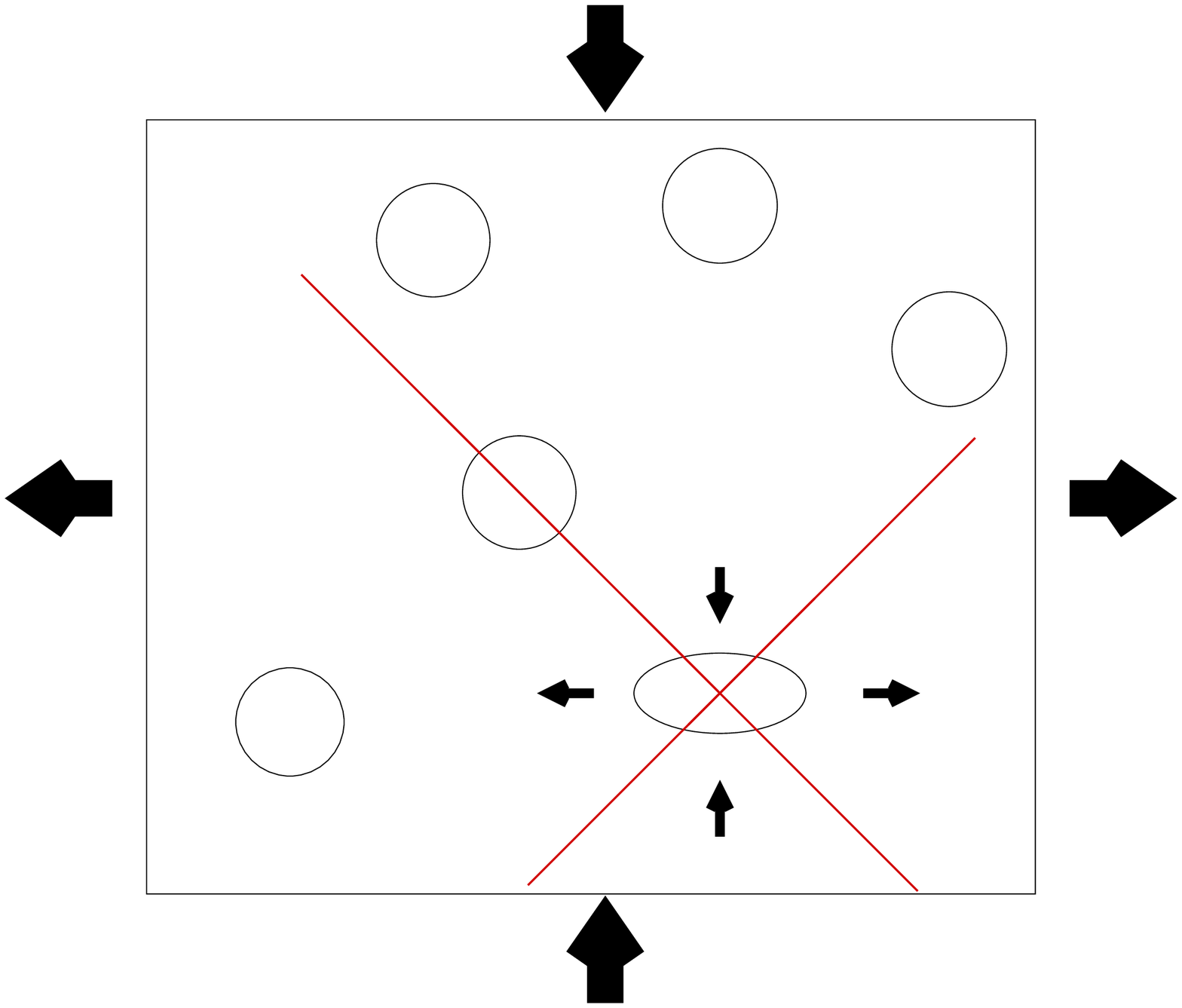}
  \includegraphics[width=.15\textwidth]{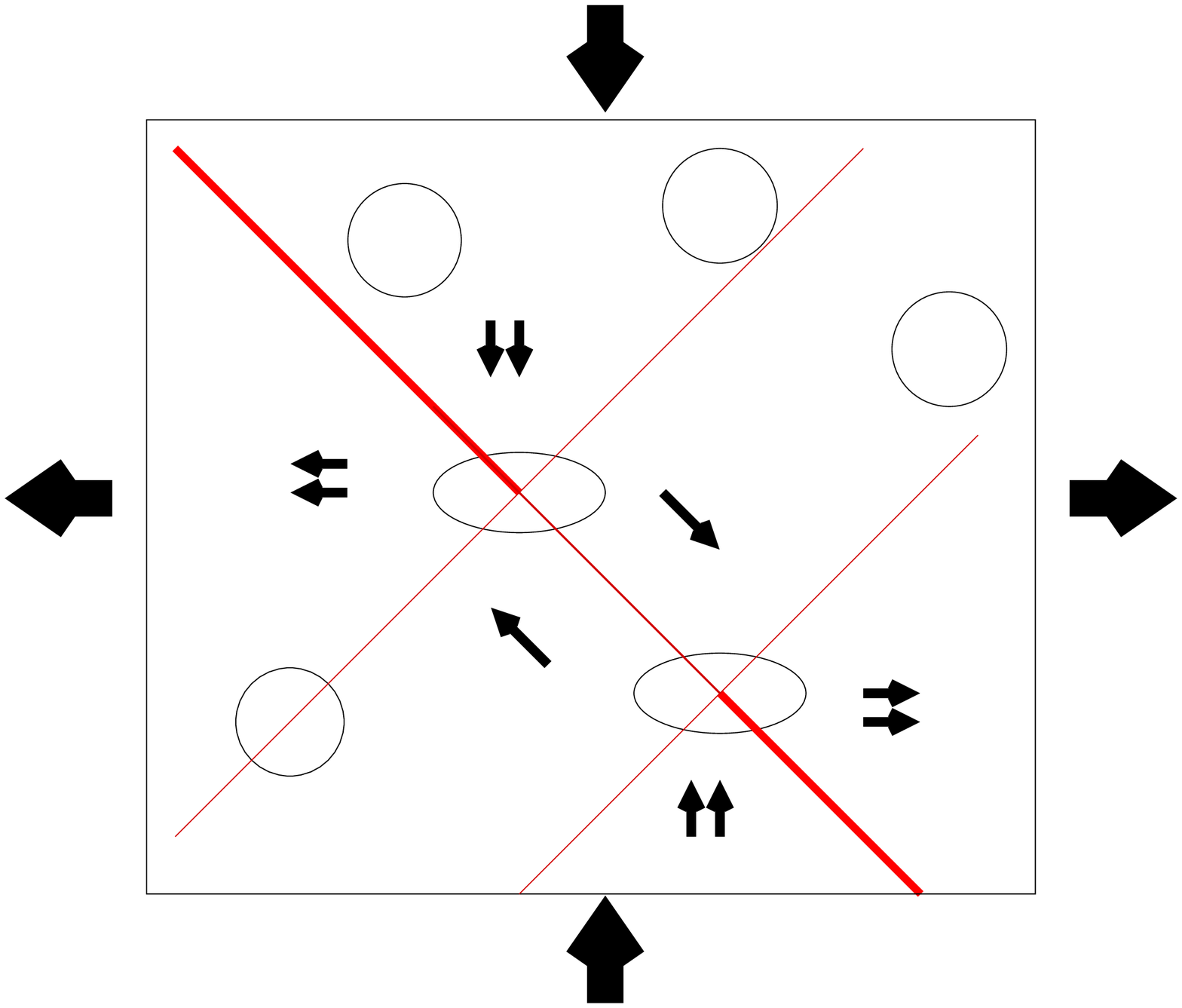}
  \caption{
Schematic representation of a banding scenario.  Sample is in tension along $y=0$ and compression along $x=0$.  Stage 1) Virgin material.  Stage 2) a) transforms and increases shear stress along red lines, increasing probability for b) to transform.  Stage 3) Should the transformation at a) induce a transformation at b), the two stress fields are roughly additive.} 
  \label{fig:bandingCartoon}
\end{figure}

The cascade mechanism is illustrated schematically in the cartoon in figure~\ref{fig:bandingCartoon}.
In frame 1, we have a virgin material with several potential shear transformation sites indicated by open circles.
As a macroscopic strain is applied, eventually, the system will become mechanically unstable (as described in detail in the previous section), and a shear transformation will be nucleated at, say, region A.
This initial event will have an associated displacement field with a $cos(2\theta)$ symmetry, giving increased (decreased) shear stresses along the lines $y=x$ and $y=-x$ (along the lines $y=0$ and $x=0$).
The potential shear transformation sites which lie away from the initially transformed site along the lines of increased stress, having had their local stress levels increased, may themselves become mechanically unstable even without further increment of the macroscopic strain.
The process favors lines of slip generated along the lines of maximal stress (which for our geometry would be along the vertical and horizontal axes); an emergent behavior much like the glide of a pair of dislocations, yet without the presence of any topologically identifiable objects.

\subsection{A Typical Cascade}
To fix these ideas, we recall and elaborate on the single event which was discussed in reference~\cite{Maloney2004}.
The system under consideration is a 200x200 system of harmonically interacting particles, with particle mixtures and preparations as described in the previous section.
A segment of the stress vs. strain curve is shown in figure~\ref{fig:stressStrainLong}.

We examine the plastic event which occurred at $\gamma=.1631$, and presume that it is representative of the kinds of events which occur at steady state.
At $\gamma=.1631$, the system has just been strained past the edge of an elastic segment -- the local energy minimum has coalesced with a barrier, leaving only an inflection point in its wake -- 
and the system is poised to undergo a non-linear, plastic rearrangement upon energy minimization.

\begin{figure}
  \includegraphics[angle=-90,width=.45\textwidth]{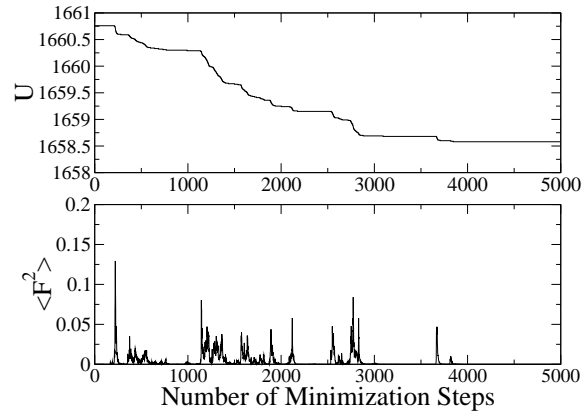}
  \caption{Energy and sum of the squares of the forces during the conjugate gradient descent for the single event at $\gamma=.1631$.}
  \label{fig:descent}
\end{figure}

The total potential energy and sum of the squares of the forces during the energy minimization are shown in figure~\ref{fig:descent}.
As we use a conjugate gradient minimization routine, we have no rigorous notion of time.
However, in steepest descent dynamics, where $dx_{i\alpha}/dt=F_{i\alpha}$, the time derivative of the energy is precisely the sum of the squares of the forces.
We have implemented a steepest descent dynamics and have checked on a single event that the resulting cascade was similar to the result of the conjugate gradient algorithm. 
This allows us to interpret the number of minimization timesteps as some measure of time. 
However, the steepest descent algorithm is by far too slow to be used systematically in this study, and we thus have to rely on conjugate gradient methods.

As the non-linear conjugate gradient algorithm progresses, during the first 100 or so minimization steps, the system behaves much as if it was making a small correction during a purely elastic relaxation, operating in an essentially linear regime. 
The energy relaxes toward a plateau, and the forces also decrease (not visible on the scale of the figure) accounting for the non-affine linear elastic corrections (discussed in detail in the previous section) necessary to return the system, roughly speaking, to where the minimum ``ought to'' be.
The forces are very small, and the energy plateau is very flat.
It is almost as if the system is at a mechanical equilibrium state... a ``quasi-basin''.
Soon, however, at about the 200-th minimization step, the forces get large and the energy drops rapidly; the system exits this quasi-basin. 

The force curve is intermittent with clusters of sharp peaks separated by quiescent periods, however it is difficult to make precise distinctions; peaks may overlap and quiescent periods may be disturbed by small rumblings down by an order of magnitude from the peaks.
\begin{figure}
  \includegraphics[width=.45\textwidth]{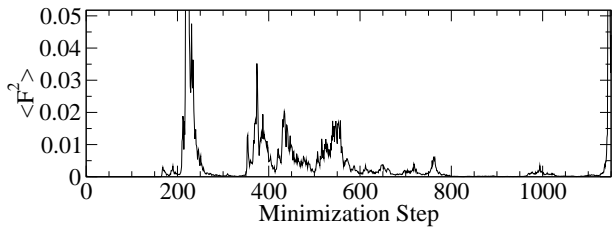}
  \caption{
Close-up of the data shown in figure~\ref{fig:descent}b during the first $1150$ minimization steps.}
  \label{fig:descentCloseup}
\end{figure}
We will focus on the cluster of force peaks which occur before minimization step $1150$ and replot this segment of the descent on a blown up scale in figure~\ref{fig:descentCloseup}.
Note the reasonably well defined initial force peak around step 230 followed by several subsequent smaller peaks up to step 800.
After step 800, there is a period of relative quiescence which which lasts through step 1100.

\begin{figure*}
\includegraphics[width=.9\textwidth]{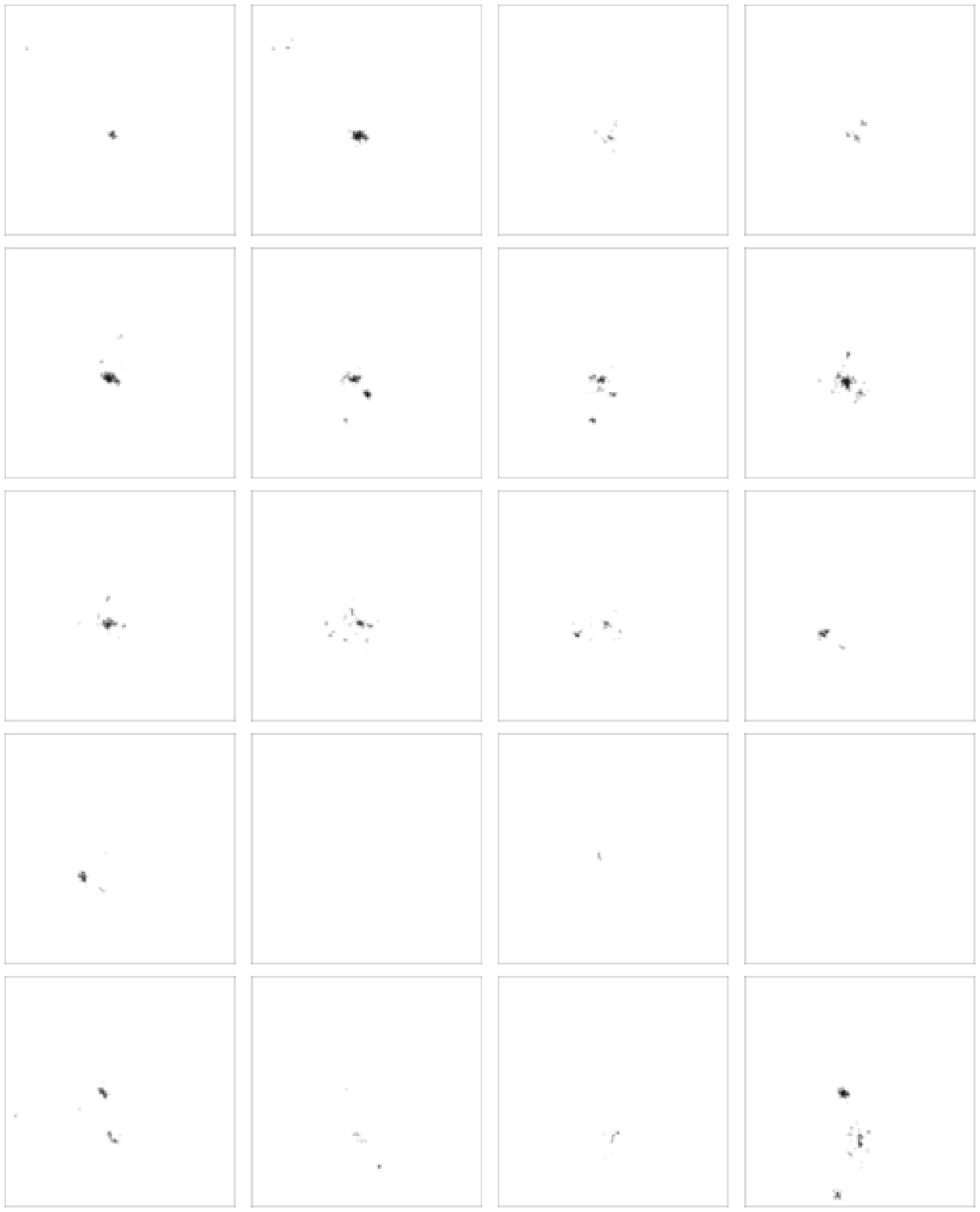}
\caption{
Incremental slip (as defined in the text) at 50 step intervals during the minimization routine.
Sequence starts at step 200 and ends at step 1150.
A particle is shaded if it slips by more than $10^{-3}$.
}
\label{fig:slipRealSpace}
\end{figure*}
\begin{figure*}
\includegraphics[width=.9\textwidth]{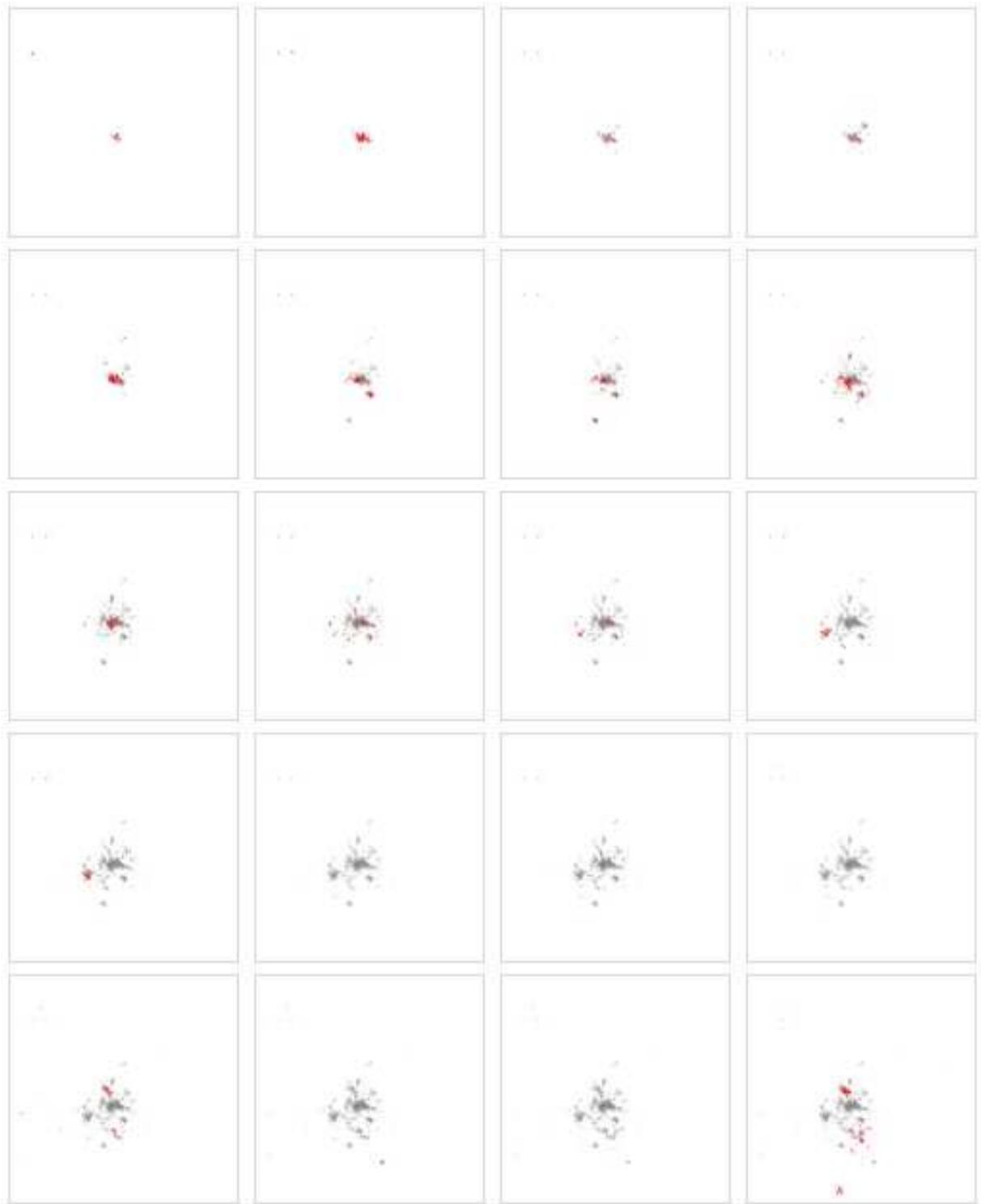}
\caption{
Cumulative slip (incremental slip superimposed in red) at the same intervals as in figure~\ref{fig:slipRealSpace}.
}
\label{fig:slipCompositeRealSpace}
\end{figure*}
\begin{figure*}
\includegraphics[width=.9\textwidth]{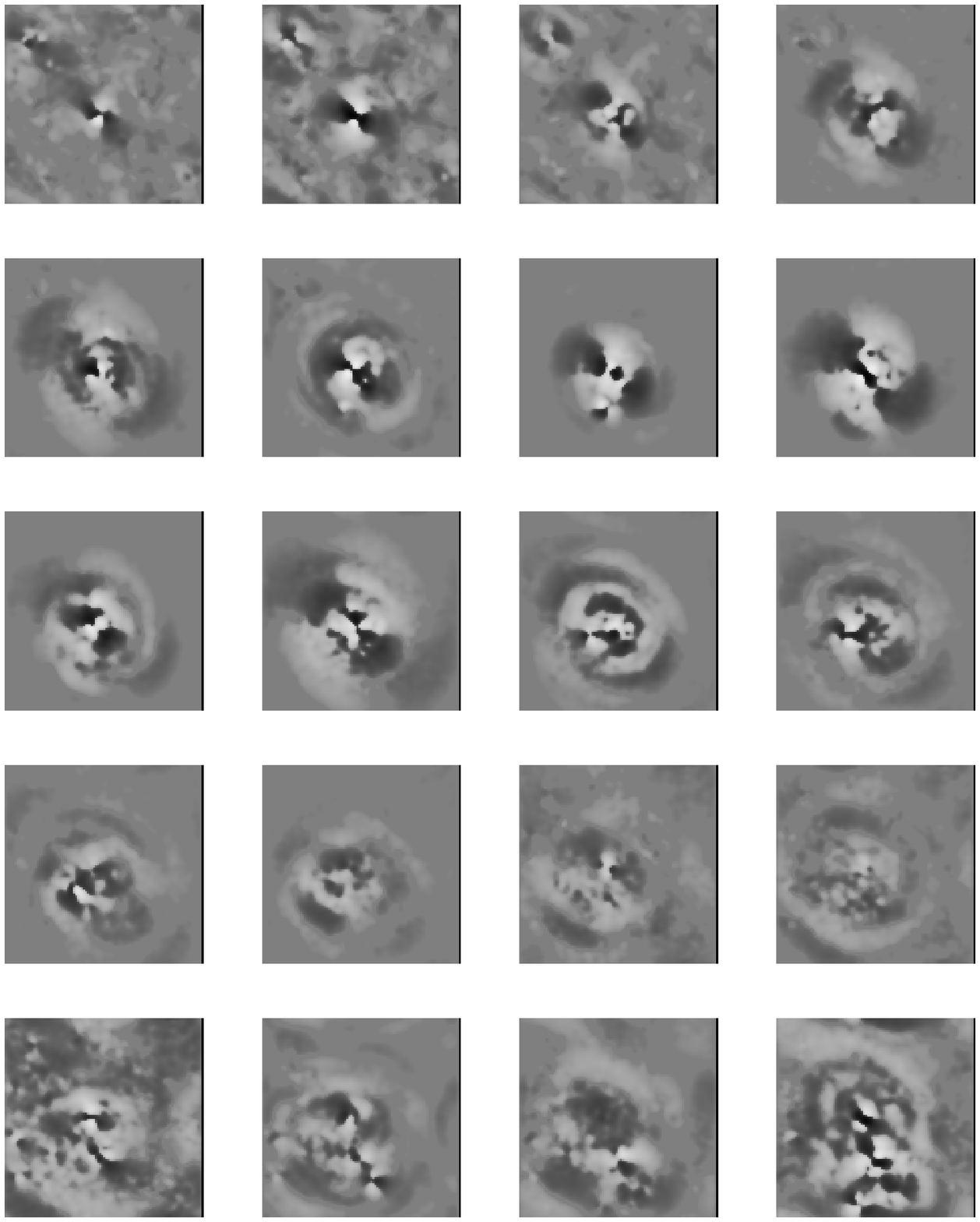}
\caption{
Local energy dissipation during the same time intervals as in figures~\ref{fig:slipRealSpace} and~\ref{fig:slipCompositeRealSpace}.
Data is on a log scale which ranges from $10^{-6}$ to $10^{1}$, with grey indicating an energy change of less than $10^{-6}$ in magnitude, white indicating an energy increase, and black, an energy decrease.
}
\label{fig:energyDisipRealSpace}
\end{figure*}

The energy descent is broken into intervals of $50$ steps each, starting at $200$ and ending at $1150$.
The slip and energy dissipation in realspace which occurs during these periods is shown in the sequence of images in figures~\ref{fig:slipRealSpace},~\ref{fig:slipCompositeRealSpace}, and~\ref{fig:energyDisipRealSpace}.
For any particular particle, we define the slip as the difference between the displacement of that particle and the average displacement of its neighbors (particles with which it is in contact); in this sense it is analogous to a discrete Laplace operation on the displacement field.
All real space structures are transformed back to the rectilinear frame by an inverse simple shear of magnitude equal to the current strain, $.1631$, such that in the plots shown, the point $(x,y)$ may be identified with the points $(x+aL,y+bL)$ where $a$ and $b$ are arbitrary integers.

The incremental slip occurring in each window of 50 minimization steps is shown in figure~\ref{fig:slipRealSpace}, and the cumulative slip (with incremental slip superimposed in red) is shown in figure~\ref{fig:slipCompositeRealSpace}.
Any particle is shaded if it has slipped by more than $10^{-3}$.
The local incremental energy dissipation is plotted, in figure~\ref{fig:energyDisipRealSpace}, on a log scale which ranges from $10^{-6}$ to $10^{1}$, with grey indicating an energy change of less than $10^{-6}$ in magnitude, white indicating an energy increase, and black, an energy decrease.

The pattern which emerges is that each of the peaks in the squared force in figure~\ref{fig:descentCloseup} corresponds to a well localized cluster of particles which undergoes large slipping relative to its neighbors.
As can be seen in figures~\ref{fig:slipRealSpace} and~\ref{fig:slipCompositeRealSpace}, the new slippage does not occur precisely on top of the slippage which occurred during previous peaks, but instead, new slippage tends to occur at the extremities of the region which has already undergone appreciable slip. 
Furthermore, the local energy dissipation which occurs concomitantly with each force peak and cluster of slipping particles can be seen to take the form of a quadrupole which is centered over the cluster of slipping particles. 
Note that the quadrupoles predominantly have energy increases (white) along the tensile axis, $y=x$, and energy decrease (black) along the compressive axis, $y=-x$.
This spatial organization can be understood in terms of the banding argument outlined above.

\begin{figure*}
  \includegraphics[width=.9\textwidth]{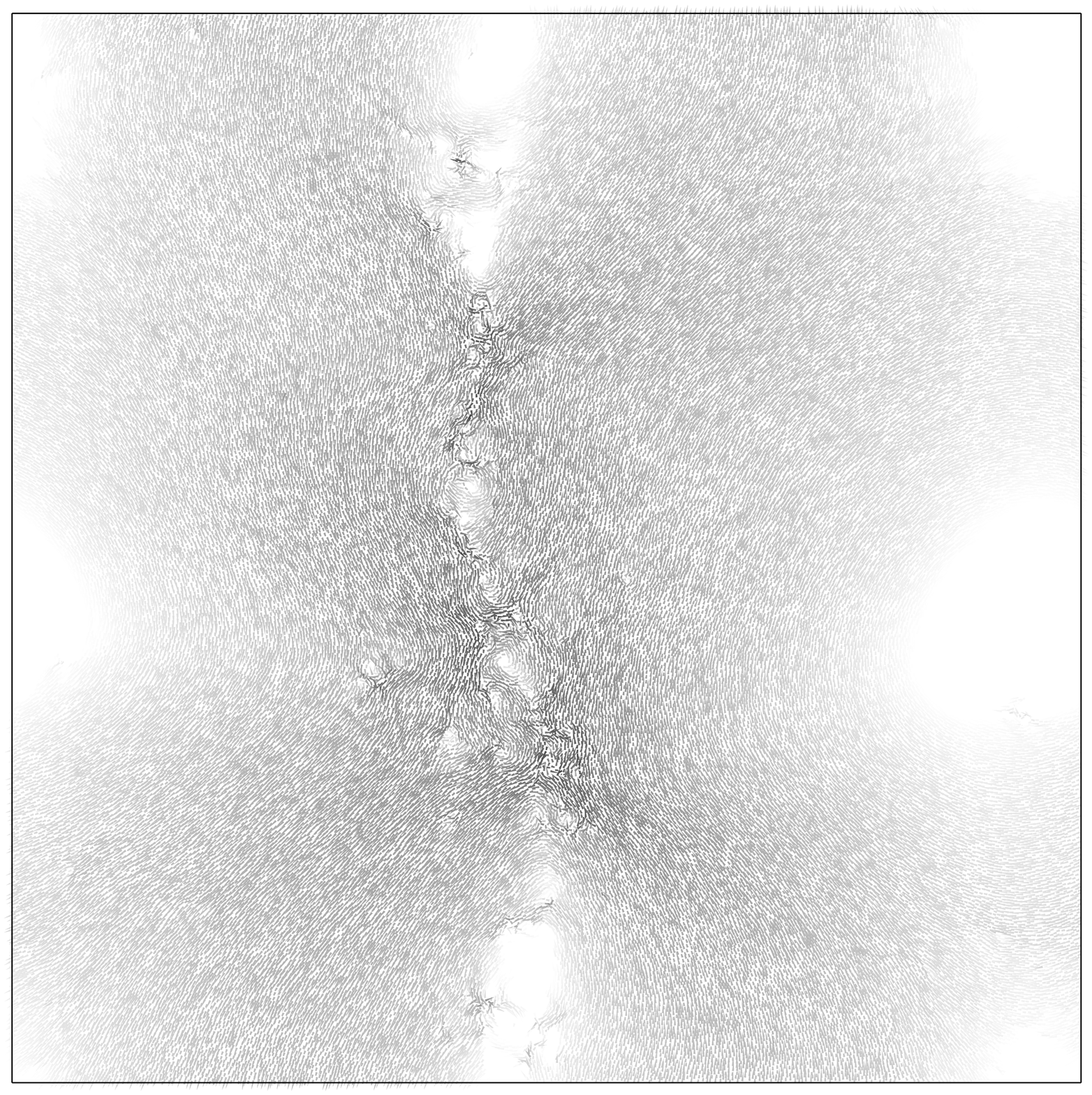}
  \caption{
    Particle displacements which occur during the entire plastic event.
    Individual arrows have a uniform length of $.5$ and a shading which is linear in the amplitude of the displacement.  The reader is encouraged to utilized the zooming features of the pdf document format to explore the fine scale structure of the displacement field.
  }
  \label{fig:dispWhole}
\end{figure*}

\begin{figure*}
  \includegraphics[width=.95\textwidth]{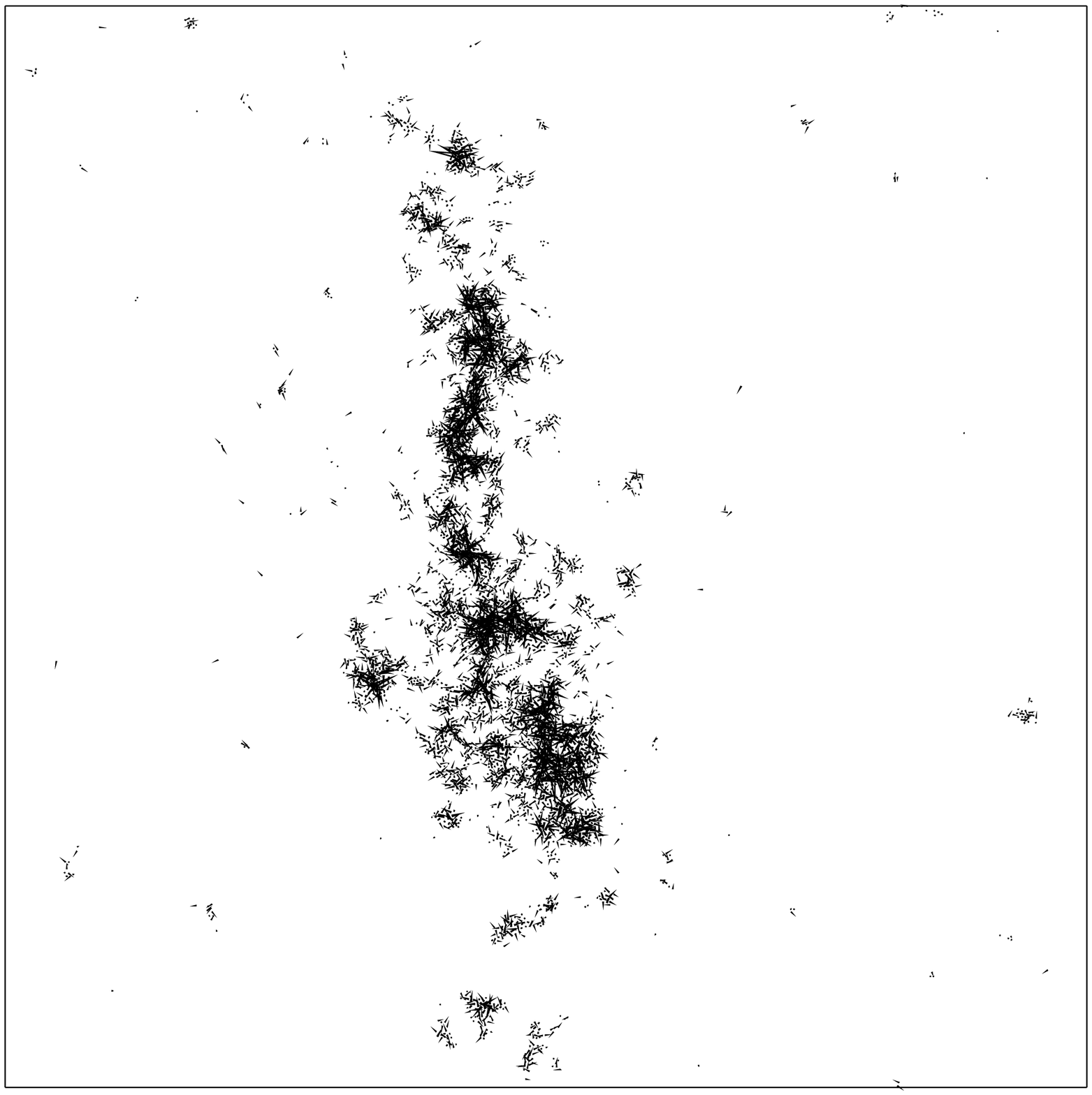}
  \caption{
    Local slip (as defined in the text) which occurs during the entire plastic event.
    Arrow lengths are equal to the magnitude of the particular slip scaled by a factor of $10$.
    Slips of amplitude less than $10^{-3}$ are not shown for clarity.
  }
  \label{fig:slipWhole}
\end{figure*}

Figure~\ref{fig:dispWhole} shows the displacement and slip which occurs after the energy relaxation is fully complete.
The displacement field is plotted such that an arrow of length $.5$ is drawn at each particle in the direction of its displacement.
The shade of the arrows represents the amplitude of the displacement on a linear scale.
This representation allows for better appreciation of the orientation of the displacement field even when its magnitude becomes small. 
Recall from the definition above that the slip is essentially a type of discrete derivative of the displacement.
We prefer to deal directly with displacements rather than energies or stresses, as the latter are essentially spatial derivatives of the former and thus much noisier.

The coherent, elastic-like behavior of the displacement field is striking.
It is consistent with the banding mechanism described above, and is roughly equivalent to the displacement field resulting from the gliding apart of a pair of dislocations or a shear crack running vertically through the system.
The cascade appears to arrest before it has spanned the system.
There is a large, weak vortex (clockwise) at about $(x=0,y=L/2)$ between the slip line and its periodic images and a large, weak hyperbolic flow at about $(x=0,y=0)$.
They are to be expected from a periodic array of incomplete vertical slip lines.
If the slip line had been complete, the resulting pattern would have likely been the text book example of a pure shear with displacements in the y direction, a gradient along the x direction, translationally invariant along the y direction, and with a discontinuity along the slip line.

Along the vertical slip itself, one sees small vortex-like (counter-clockwise) displacement fields in which the magnitude of the displacements is markedly smaller than the displacements of particles which slip.
These smaller vortices represent regions of the material along the slip line which have \emph{failed} to slip and can be thought of as resulting from a uniform line of elastic quadrupoles (a slip line) with a gap in the line.  
Not surprisingly, these vortex-like regions of ``unbroken'' material which appear in the displacement field in figure~\ref{fig:dispWhole} correspond to regions which are absent from the plot of the slip field in figure~\ref{fig:slipWhole}.

\subsection{Discussion}
In this section, we have examined one particular, typical (the magnitude of stress relaxation was around the average, and the event was taken at random from the set of all plastic events) plastic event in a 200x200 system of harmonically interacting particles and shown that the behavior was consistent with the picture of a cascade of elastically interacting localized plastic yielding events organized into a line of slip along the vertical bravais axis of the simulation cell. 
We have observed other such events aligned along either the vertical or horizontal axis of the simulation cell.
Such organization of local slippage events into lines of slip was observed long ago in experiments on bubble rafts by Argon and Kuo~\cite{Argon1979a} and, more recently, in confined films by Abd el Kader and Earnshaw~\cite{el1999}.
Although experimental observations of these features are limited to studies of soap bubbles, observation in numerical simulations are quite rich.

In their vertex model for dry foams, Okuzono and Kawasaki~\cite{Okuzono1995} observed that in the limit of small strain rate, their system underwent large events, and the authors proposed an analogy with avalanches in sandpile models.
The displacement fields associated with these events were similar to the ones we have shown here, with the system exhibiting two ``elastic-like'' regions slipping with respect to each other along a line at 45 degrees to the principle axes of applied strain (\emph{c.f.} figure 5b in reference~\cite{Okuzono1995}). 
We do not know of any attempts to observe such displacement fields in models of wet foams such as Durian's bubble model~\cite{Durian1995,Durian1997}.

Evidence of such transient, slip bands is also observed in the simulations of compressed granular materials by Aharonov and Sparks~\cite{Aharonov2002} and Kuhn~\cite{Kuhn1999,Kuhn2003}.
These simulations took into account the Coulombic friction between particles at contact and were performed at finite, but small strain rates, yet the emergent behavior seems, at least qualitatively, insensitive to these details.
Note, however, that one might expect qualitatively different behavior as the truly rigid, hard sphere limit is approached.

Simulations were performed by Leonforte and co-workers~\cite{LeonfortePrivate} which were similar to the ones we have described here but with rigid walls at $y=0$ and $y=L$ rather than fully periodic boundary conditions.
Not surprisingly, the vertical bands are suppressed, and transient horizontal bands emerge.
\section{Finite Size Scalings}
\label{sec:scaling}
The cascades discussed in the previous section which comprise the plastic events will now be shown to give rise to simple scalings of various measures of the event size with the length of the simulation cell.
In the following, we classify each strain step taken in the simulation as either elastic or plastic: if the stress was positive, a strain step was considered to be plastic if it resulted in an energy decrease; if the stress was negative, a step was considered to be plastic if it resulted in an energy increase
~\footnote{This case occurs only in the smallest systems, and even then it is infrequent.}; 
otherwise, it was considered to be elastic.
We suppose that this operational definition corresponds closely to the rigorous definition of a plastic event in terms of the onset of a catastrophe via the vanishing of an eigenvalue of the hessian matrix, but as we emphasized above, care must be taken to ensure that we take small enough strain steps such that we remain in the quasistatic regime.
Note however that, in principle, there may always be small plastic discontinuities whose energy drop would be masked by the elastic energy increase for a given strain step size.
All the data used for the statistical analysis were gathered from an ensemble of 8 systems at strains between 1 and 2 for each box size and interaction potential.

\begin{figure}
\includegraphics[width=.48\textwidth]{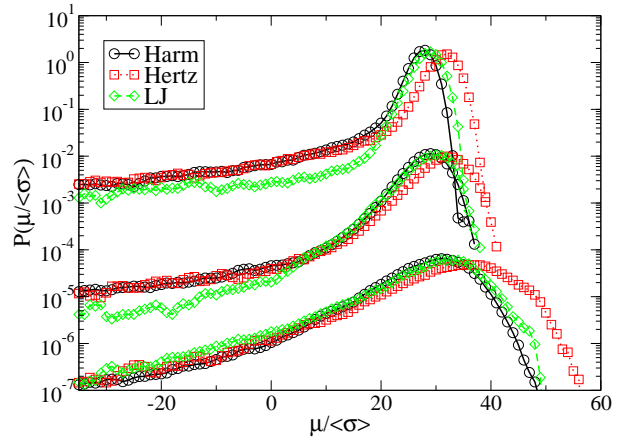}
\caption{Distribution of instantaneous modulus (defined by the finite difference, $\Delta\sigma/\Delta\gamma$) normalized by the average stress during steady flow.
Each group of three curves corresponds to a particular system length (12.5,25,50) arbitrarily shifted vertically for clarity with increasingly longer lengths shifted upward.
Legend is as follows: circles with solid line (black): Harmonic; squares with dotted line (red): Hertzian; diamonds with dashed line (green): Lennard-Jones.
}
\label{fig:distMu}
\end{figure}

Before we begin our discussion of the statistics of the plastic events, we first report, in figure~\ref{fig:distMu}, on the distributions of the instantaneous shear modulus, $\mu=\frac{\Delta\sigma}{\Delta\gamma}$, for the various system sizes and interaction potentials.
We normalize the values by the average flow stress, $\langle\sigma\rangle$ and note that the distributions are essentially unchanged if we include/exclude the plastic events from the analysis.
The average flow stresses for all interactions and system sizes are reported in table~\ref{tab:stress}.
For each potential, there is a slight increase with size of the average, $\langle\sigma\rangle$, which is noted in table~\ref{tab:stress} --- this effect seems to saturate for the larger systems.

The distributions of the moduli are all essentially Gaussian in the neighborhood of the most likely value, but with tails on the soft side, indicative of the non-linear yielding upon approach to the catastrophes discussed above.
The peak occurs at a dimensionless modulus of about 28 for the Harmonic and Lennard-Jones systems and at a slightly higher value for the Hertzian system.
This implies that, generically, these dense, disordered systems have a flow stress which is about $.035$ times the characteristic modulus, regardless of the nature of the interaction potential.
This value of the ``flow strain'', $\frac{\langle\sigma\rangle}{\mu}$, is roughly consistent with simulations of various models of foams in 2D~\cite{Okuzono1993,Okuzono1995,Hutzler1995,Durian1995,Durian1997} (in the dense limit, away from the loss of rigidity) and the 3D simulations of yielding of a Lennard-Jones glass~\cite{Varnik2003,Varnik2004,Rottler2003}, although it is somewhat higher than Johnson's recently reported universal flow strain in 3D glasses~\cite{Johnson}.
It is not clear precisely how this flow strain, measured in steady flow, relates to the yield strain measured by Mason and coworkers in oscillatory rheological experiments on microemulsions~\cite{Mason1996}, but the order of magnitude of a few percent is in rough agreement.

We stress that it is our ability to resolve the elastic behavior and measure a well defined modulus which gives us confidence that we have chosen a strain step which is small enough to properly resolve the quasistatic behavior.
For larger strain steps, the well defined peak disappears, the stress essentially makes a random increase or decrease at each step, and the quasistatic behavior --- i.e. the separation of plastic from elastic events --- is lost.

\begin{table}
\begin{tabular}{|l||c|c|c|}\hline
 & 12.5 & 25.0 & 50.0 \\ \hline\hline
Harm & 6.5E-2 & 6.9E-2 & 7.2E-2 \\ \hline
Hertz & 5.2E-3 & 6.0E-3 & 6.0E-3 \\ \hline
LJ & 1.5E0 & 1.7E0 & 1.8E0 \\\hline
\end{tabular}
\caption{
Average stress during steady flow for the three different system sizes and interaction potentials.
}
\label{tab:stress}
\end{table}

For the purposes of this work, the more interesting quantities are the various measures of cascade size.
Obvious choices are the distributions of the stress drops, the energy drops, and the lengths of the elastic segments.
As the patterns we observed in the 200x200 system of harmonic discs were clearly 1D features, one might expect that the distributions of the various quantities which characterize the event size would be invariant when properly rescaled by the length of the box.

First consider the distribution of stress drops.
Since the events we observe are predominantly organized into lines of slip which extend across the length of the cell, we expect that such an event should release an amount of stress equal to $\Delta\sigma\sim\mu\delta\epsilon\sim\mu (a/L)$ where $a$ is some measure of the amplitude of the slip at the site of the cascade and $L$ is the length of the box.
Since the system is in steady state, the stress built up in the strain interval, $\mu\Delta\gamma$,  between these events must be equal, on average, to the stress released in an event, $\Delta\sigma$, so we must have that, $\Delta\gamma\sim a/L$.
The stress drops can be related to the energy drops if we make the simple assumption that, on average, the energy release is elastic; that is:
\[\Delta U\sim L^2 \frac{\langle\sigma\rangle\Delta\sigma}{\mu}\sim aL\langle\sigma\rangle\]

Now that we have considered the energetics of the plastic events, we show how the same size effects dictate their relative frequency.
During the elastic segments, the stress increase is $\Delta\sigma_{\text{el}}\sim \mu\Delta\gamma$.
In steady state, this elastic stress increase must be balanced, on average, by the plastic dissipation, so we simply have:
\[\Delta\gamma\sim \frac{\Delta\sigma}{\mu}\sim \frac{a^2\lambda}{L}.\]
So we see that the relative frequency of the plastic events should not even depend on the energy scale of the underlying interaction potential.

\begin{figure}
\includegraphics[width=.4\textwidth]{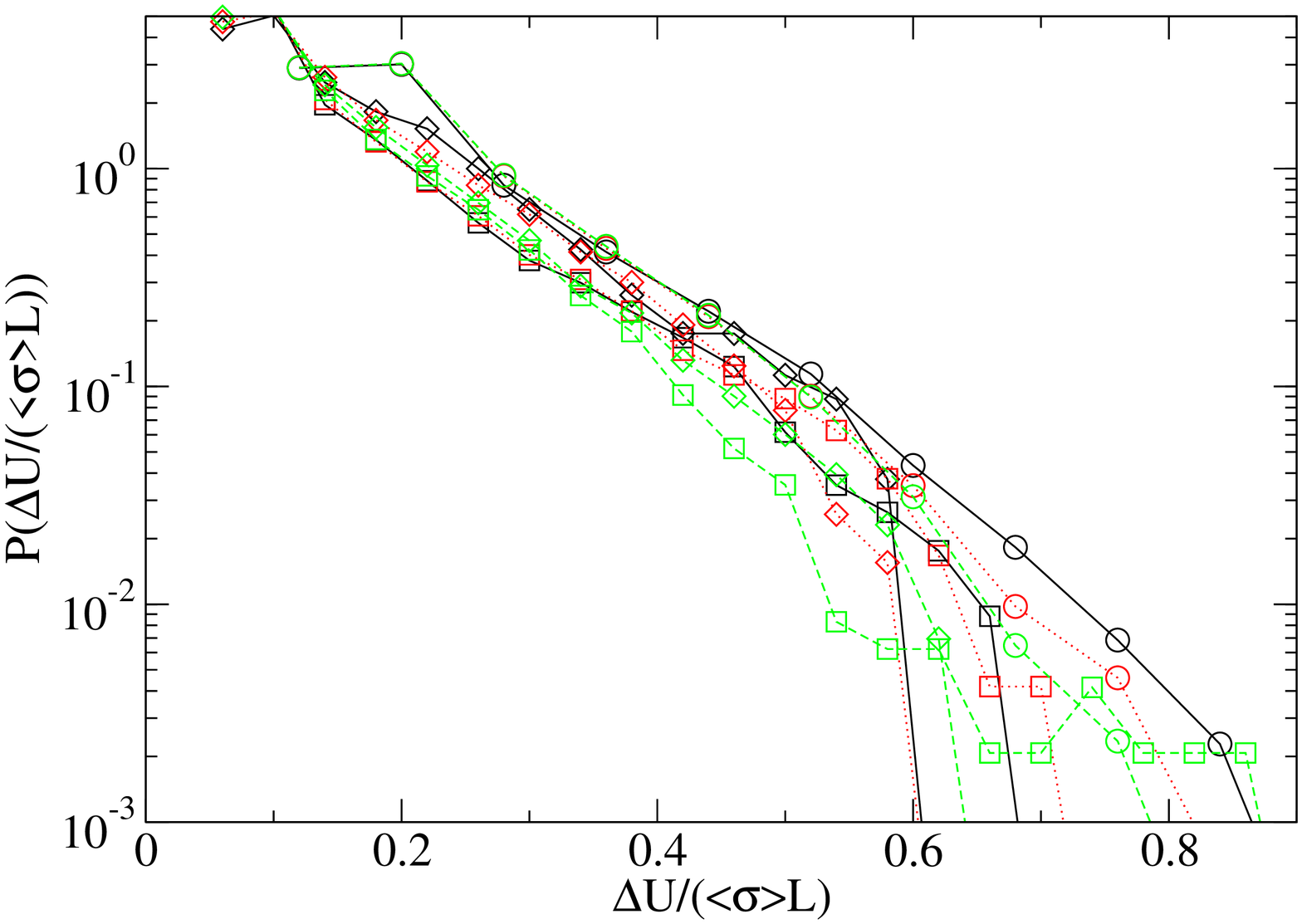}
\includegraphics[width=.4\textwidth]{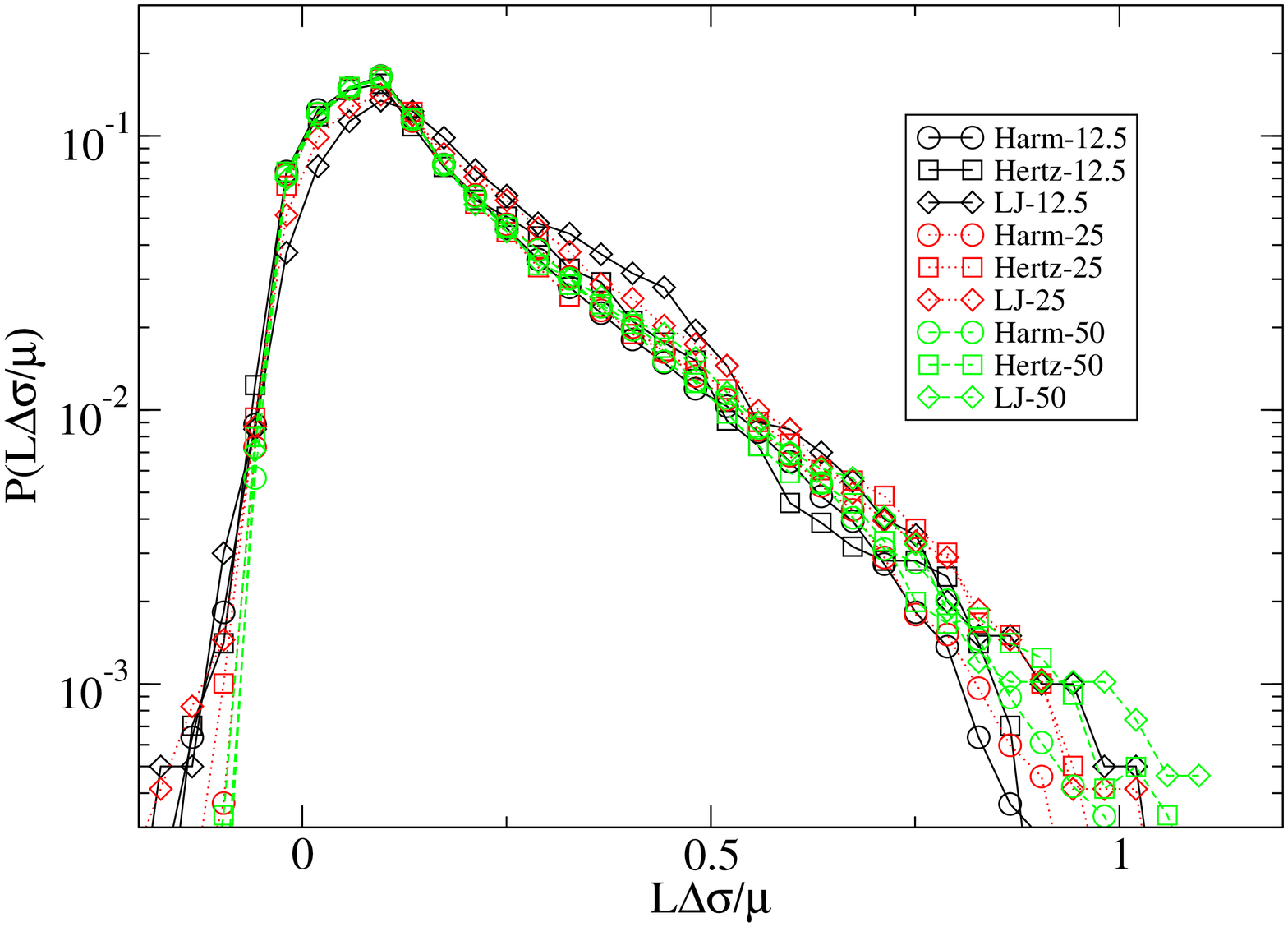}
\includegraphics[width=.4\textwidth]{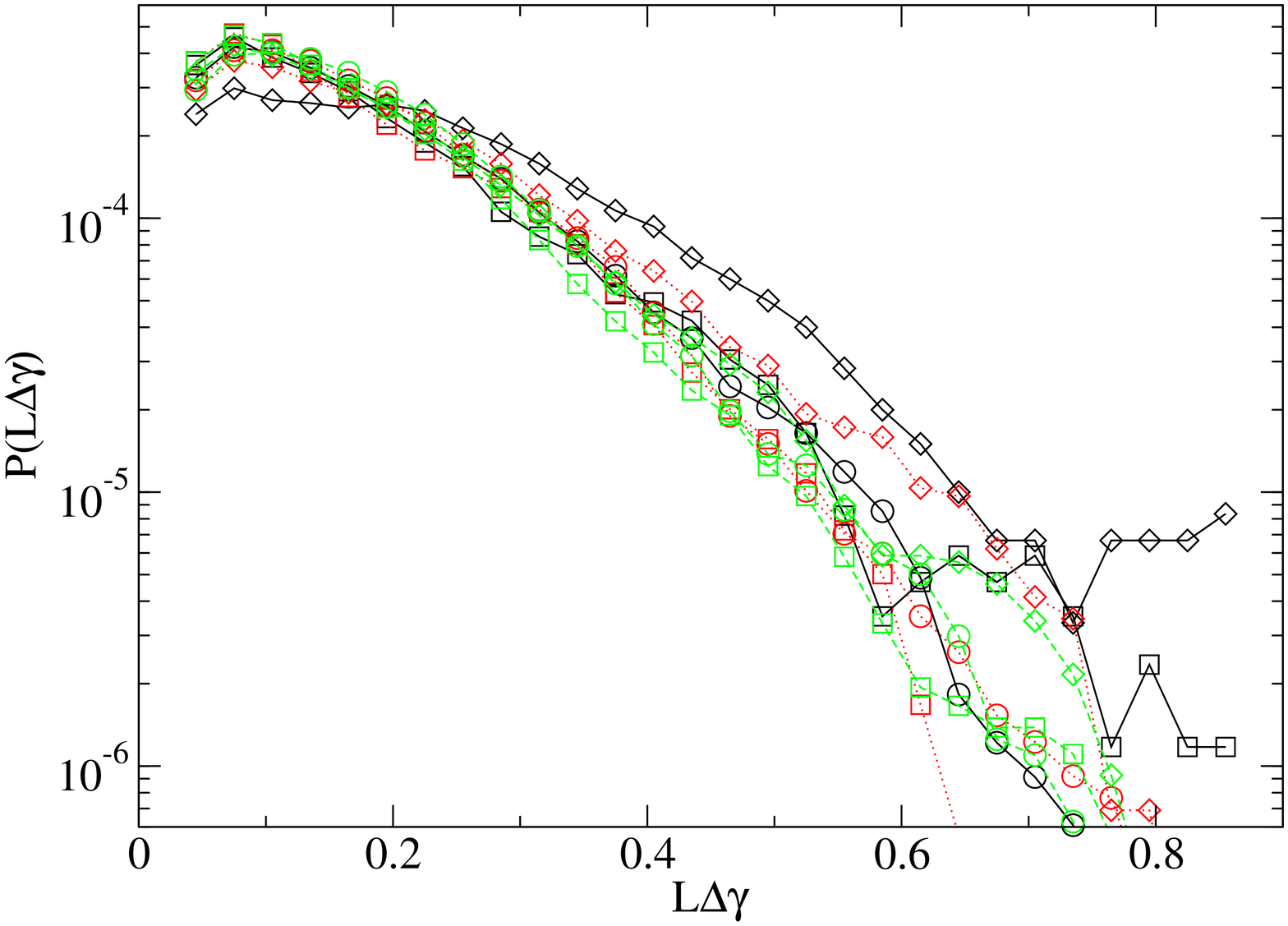}
\caption{Distribution of (top to bottom): $\frac{\Delta U}{L\langle\sigma\rangle}$, $\frac{L\Delta\sigma}{\mu}$, and $L\Delta\gamma$.
Where $\Delta U$ is the energy drop for a plastic event, $\Delta\sigma$ is the stress drop for a plastic event, and $\Delta\gamma$ is the length of an elastic segment.
Legend is as follows: solid (black): $L=12.5$; dotted (red): $L=25$; dashed (green): $L=50$; circles: Harmonic; squares: Hertzian; diamonds: Lennard-Jones.
}
\label{fig:allDropScaled}
\end{figure}

\begin{figure}
\includegraphics[width=.48\textwidth]{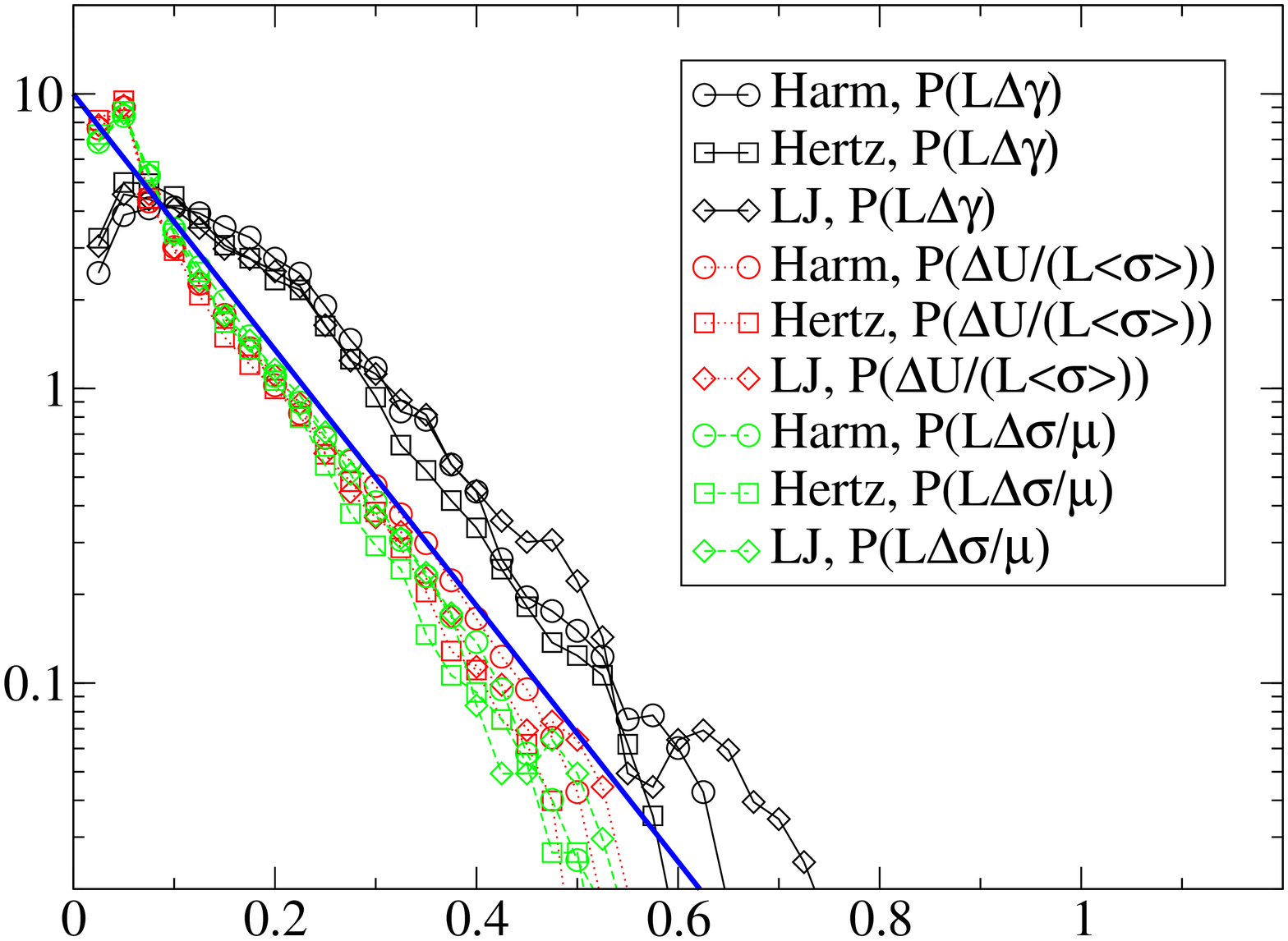}
\caption{Probability distributions of $\frac{\Delta U}{L\langle\sigma\rangle}$, $\frac{L\Delta\sigma}{\mu}$, and $L\Delta\gamma$ for all three interaction potentials.
Only the 9 curves corresponding to the largest systems are shown for clarity.
}  
\label{fig:allDropScaledOverlay}
\end{figure}

We now proceed to plot the various distributions: $P(\frac{\Delta U}{L\langle\sigma\rangle}), P(\Delta\sigma(\frac{L}{\mu}))$, and $P(L\Delta\gamma)$, for all three system sizes and all three interaction potentials.
All 27 curves collapse onto a single master curve which is very well fit by an exponential (with better collapse for the larger system sizes).
For clarity, in figure~\ref{fig:allDropScaled} we first show the distribution for each of the three quantities --- energy drop, stress drop, strain interval --- in its own plot for all 9 systems, then, in figure~\ref{fig:allDropScaledOverlay}, show the 9 curves corresponding to only the largest system size, but for all three quantities and for all three interaction potentials.
This characteristic value for $a$ which one extracts from the master curve is a few tenths of a particle diameter, in good agreement with the discontinuity in the displacement fields shown above.
The precision of the collapse in figure~\ref{fig:allDropScaledOverlay} is quite striking and indicates that the nature of the slip which occurs along the cascade line has little to do with the precise nature of the interaction potential.

The event rate scalings, in particular, have profound consequences concerning both fundamental and technical issues.
For any arbitrarily small strain step size, there will always be systems large enough, such that the step size is no longer small enough to resolve the elastic behavior, many independent plastic events will be simultaneously nucleated at every strain step, and the quasistatic behavior will break down.
Thus, it is technically important, for any quasistatic simulation, to verify that one is properly resolving the elastic behavior, \emph{independently for each system size}!
For the present study, this is evidenced by the peak-like behavior in figure~\ref{fig:distMu}.
We note that, had the elementary relaxation events been uncorrelated, the plastic event rate would have scaled extensively, as $L^2$, so the correlations reduce the frequency of plastic events relative to what would have been observed for independent uncorrelated events. 
\section{Conclusion}
\label{sec:conclusion}
In conclusions, we have shown that elementary plastic events take the from of cascades of shear transformations in the athermal quasistatic limit for a general class of densely packed simple amorphous materials.  
The incremental stress and displacement fields associated with the cascades are quite reminiscent of micro-structural shear cracks which immediately heal themselves after cracking.
These local shear transformations might be roughly thought of as the elementary objects responsible for plasticity in amorphous material, analogous to the dislocations which are thought to be responsible for plasticity in crystals.

The locus of rearrangement shows some spatial correlation from one event to the next, but these correlations decay quickly after just a few plastic events.
We observe no evidence for the kinds of pronounced \emph{persistent} shear localization which is seen in many experiments and simulations of sheared amorphous materials where hard walls are employed to drive the system, and we find it likely that the \emph{persistent} localization observed elsewhere is due largely to effects of the boundary.
Future investigations into the role of the boundaries will be crucial in both atomistic studies~\cite{Fu2001,Varnik2003,Rottler2003,Xu2005} and mesoscopic models~\cite{Picard2004,Picard2005}.

The detailed picture of the onset of individual plastic events which we developed showed that directions of low curvature on the potential energy surface, when driven to zero by the strain, are responsible for nucleating the cascades of shear transformations.
These low-lying modes which were shown to have an essentially ``shear-transformation-like'' quadrupolar character, were found to play a dominant role in the non-affine elastic displacement fields, even reasonably far away from the onset of a cascade, and might be observed experimentally in this way.
Strikingly, at least in the systems studied here, these modes are not observable via looking at local stresses (e.g. as in reference~\cite{Srolovitz1983}), or local Born values of the elastic constants, and can only be observed
through the non-affine elastic response which is a collective property of the potential energy surface.

We showed that the organization of the shear transformations into cascades during the individual plastic events caused a scaling of the average event size with the length of the simulation cell regardless of the underlying interaction potential.
These strongly correlated events, which involve lengths as large as the simulation cell, are in qualitative agreement with various numerical simulations, both atomistic~\cite{Yamamoto1997,Yamamoto1998} and mesoscopic~\cite{Bulatov1994b,Picard2005}.
They induce scalings with the length of the system for various measures of event size (energy drop, stress drop, and elastic segment length) which would have been incorrectly predicted from an underlying picture of uncorrelated localized events. 
Furthermore, various interaction potentials (Lennard-Jones, harmonic and hertzian springs) were shown to exhibit nearly identical behavior upon appropriately adjusting the energy scale for the given potential.

Although, we were able to measure the properties of the nucleating modes quite precisely (at least in moderately sized systems), once cascades were initiated, the picture became quite complicated with simultaneous and overlapping shear transformations.
In general, our work highlights the fundamental difficulties involved in decomposing any cascade into constituent elementary events.
Thus, important questions which are relevant to the construction of plasticity theories, such as the spatial structure of a complete, isolated shear transformation~\cite{Langer2001,Picard2004}, remain open.
Even so, we were able to demonstrate that the predominant activity during the cascades was qualitatively characteristic of local shear transformations, with incremental displacements and associated mechanical fields having predominantly quadrupolar character and the cumulative displacements and mechanical fields being reminiscent of narrow lines of slip.


One of the most important directions for future study will be extending to finite temperatures and strain rates.
The scaling analysis performed by Yamamoto and Onuki~\cite{Yamamoto1997,Yamamoto1998} showed that in sheared supercooled liquids, the dynamical correlation length becomes temperature independent at small enough temperatures --- the ``strong-shear'' regime --- whence it scales like the strain rate to some negative power.
The recent \emph{athermal} mesoscale model of plasticity proposed by Picard and co-workers~\cite{Picard2005} also exhibits a diverging length, but with a stronger divergence than that reported in the 2D simulations by of Yamamoto and Onuki. 
Recent work by Langer, Liu and co-workers and Berthier and Barrat has suggested that imposed strain rate induces an \emph{effective temperature} in athermal systems~\cite{Langer2000,Ono2002}, much the same as the effective temperature in unsheared supercooled liquids~\cite{Barrat2001a,Berthier2002a,Berthier2002} defined in terms of effective fluctuation dissipation relations~\cite{Cugliandolo1997}, and consistent with the picture of Yamamoto and Onuki of decreasing correlations upon increasing either strain rate or temperature.

This raises important questions regarding the correlated motions in both the atomistic and mesoscopic simulations.
What role do shear transformations play at finite temperature and strain rate both in the strong-shear and thermal (aging) regimes, and how robust with respect to finite drive is the cascade mechanism?
Is the underlying mechanism for the heterogeneity in the strong-shear regime different than the thermal regime?
Does the strong-shear scaling relation, $\xi\sim\dot{\gamma}^{-1/4}$, of Yamamoto and Onuki continue to hold below the glass transition, or is there a crossover to the $\xi\sim\dot{\gamma}^{-1/2}$ behavior observed in the mesoscale model of Picard \emph{et. al.}?
More generally, is there any difference between a supercooled liquid in the strong-shear regime and an amorphous solid?
These are some of the fundamental questions which must be addressed in order to make progress` toward a coherent theory of sheared amorphous material.
\\

\acknowledgments
We would like to thank Carlos Garcia-Cerva and Tomas Opplestrup for useful discussions regarding continuous minimization algorithms.  
We thank Jean Carlson, Christiane Caroli, Michael Dennin and Anthony Foglia for many useful discussions and critiques throughout the course of this work.
CM received support through the LLNL University Relations Program under the auspices of the U.S. Department of Energy by the University of California, Lawrence Livermore National Laboratory under Contract No. W-7405-Eng-48.
AL was supported by the NSF under grants DMR00-80034 and DMR-9813752, by
the W. M. Keck Foundation, and EPRI/DoD through the Program on Interactive Complex Networks. 
CM would also like to thank J.S.~Langer and V.V.~Bulatov for guidance and support and the MSTD/HPCMS group for its hospitality.
\bibliography{../../masterBib/junk}
\end{document}